\documentclass[preprint,12pt]{elsarticle}

\usepackage{subfloat}
\usepackage{amssymb}
\usepackage{amsmath}
\usepackage[mathscr]{euscript}
\usepackage{dcolumn}
\usepackage{bm}
\usepackage{psfrag}
\usepackage{lipsum}
\usepackage{caption}
\usepackage{xcolor}
\usepackage[colorlinks=true,linkcolor=blue,citecolor=blue,urlcolor=blue]{hyperref}
\usepackage{cancel}
\usepackage{lineno}
\usepackage{float}
\usepackage{longtable}
\usepackage{booktabs}
\usepackage{multirow}
\usepackage{comment}
\usepackage{placeins}
\usepackage{tabularx}   

\journal{Arxiv}

\begin{document}

\begin{frontmatter}

\title{Supervised machine learning of compressible flow past a rotating cylinder} 

\author[inst1]{Sanjeev Kumar}

\author[inst1]{Santosh Kumar}

\author[inst1]{Aditi Sengupta}
\ead{aditi@iitism.ac.in}

\affiliation[inst1]{organization={Department of Mechanical Engineering},                       addressline={Indian Institute of Technology, Dhanbad}, 
            city={Dhanbad},
            postcode={826004}, 
            state={Jharkhand},
            country={India}}

\begin{abstract}
High-fidelity numerical simulations of compressible flow past a rapidly rotating cylinder are used to investigate the evolution of aerodynamic loads and flow instability over a wide range of Reynolds numbers ($Re_{\infty}$ = 1000 - 6000). The study reveals a transition from periodic vortex shedding to complex multi-mode oscillatory states, with a critical bifurcation identified near $Re_{\infty} \approx 5650$. Spectral analysis of lift and drag signals shows the emergence and interaction of multiple dominant frequencies, accompanied by amplitude modulation and nonlinear mode coupling in the post-bifurcation regime. Notably, the maximum drag and lift coefficients exhibit strong non-monotonic variations, while the onset time of instability departs from a smooth parabolic trend beyond the critical $Re_{\infty}$. To model these highly nonlinear dependencies, data-driven approaches are systematically explored using a database of 101 high-fidelity simulations ($\approx 10^6$ core hours). Polynomial regression provides baseline fits but fails to capture localized fluctuations near bifurcation. Bayesian regression frameworks employing B-spline and Gaussian radial basis functions improve flexibility and uncertainty quantification, with spline-based models demonstrating superior performance in capturing piecewise nonlinear trends. Artificial neural networks (ANNs) are then developed as high-capacity surrogate models, achieving excellent predictive accuracy for maximum lift coefficient and instability onset time, while maintaining reasonable fidelity for the more challenging drag coefficient. Beyond regression, the ANN is further evaluated as a generative model to reconstruct flow behavior at unseen $Re_{\infty}$. A hierarchical refinement strategy is introduced, wherein the trained network predicts intermediate states in the critical $Re_{\infty}$-regime, at progressively finer resolutions. Results show that when trained on high-fidelity data, ANN-based models can serve as efficient and reliable surrogates for complex fluid dynamics problems, enabling accurate prediction and reconstruction of bifurcation-driven flow behavior at a fraction of the computational cost.
\end{abstract}



\begin{keyword}

data-driven modeling \sep high performance computing \sep bifurcation analysis \sep rotating cylinder \sep compressible flow \sep supervised learning

\end{keyword}

\end{frontmatter}

\section{Introduction}
\label{sec:1}

The flow past a rotating cylinder is a canonical problem in fluid dynamics that has attracted sustained attention due to its rich physics and wide range of engineering applications. The introduction of rotation fundamentally alters the wake structure through asymmetric boundary layer development, lift induced by Magnus-Robins effect, and modification of vortex shedding dynamics \cite{prandtl1926application}. These effects are relevant to applications such as rotating machinery, flow control devices, turbomachinery blade cooling, and aerodynamic lift enhancement systems \cite{modi1997moving}. At moderate Reynolds numbers, rotation is known to suppress or delay vortex shedding \cite{mittal2003flow, pralits2010instability}, while at higher rotation rates, it can induce complex wake transitions, including asymmetric shedding, mode switching, and even complete wake stabilization \cite{sengupta2003temporal}. Classical studies, both experimental and numerical \cite{stojkovic2002effect, glauert1957flow, nishioka1978mechanism, pralits2013three}, have extensively characterized these phenomena in the incompressible regime, focusing on parameters such as lift and drag coefficients, Strouhal number variation, and wake topology.

Despite these advances, most prior investigations have been restricted to incompressible formulations, where density variations are negligible. However, in many practical scenarios, particularly in high-speed rotating systems, turbomachinery \cite{sengupta2020effects, sengupta2024separation}, and compressible external flows \cite{sengupta2024thermal}, the assumption of incompressibility becomes inadequate. Compressibility effects introduce additional physical mechanisms, including density–pressure coupling, acoustic wave propagation, and modified instability characteristics. These can significantly alter vortex shedding frequency, force coefficients, and transition behavior \cite{salimipour2019surface, sundaram2021multiscale, sengupta2023compressibility}. Furthermore, at high rotation rates, local Mach numbers in the vicinity of the cylinder can become non-negligible even if the freestream Mach number is modest, necessitating a fully compressible formulation to accurately capture the flow physics \cite{kumar2011flow, suman2022novel}.

A limited number of studies have addressed compressible flow past rotating cylinders \cite{salimipour2019surface, teymourtash2017compressibility, sundaram2021multiscale}, primarily focusing on low-to-moderate Mach numbers and moderate rotation rates. These works have highlighted the influence of compressibility on lift enhancement, drag reduction, and wake asymmetry, as well as shifts in dominant shedding frequencies. However, systematic investigations at high rotation rates combined with compressible effects remain scarce \cite{sengupta2025bifurcation}, particularly in the context of transitional and post-bifurcation flow regimes. The interplay between compressibility, rotation-induced shear, and nonlinear wake dynamics in such regimes is not yet fully understood, and there exists a clear gap in the literature regarding high-fidelity characterization of these flows across a broad range of Reynolds numbers ($Re_{\infty}$).

In parallel with advances in computational fluid dynamics, supervised machine learning techniques have emerged as powerful tools for modeling complex nonlinear systems. In the context of bluff body flows, several studies have employed regression models \cite{fukami2020assessment}, reduced-order models \cite{hasegawa2020machine, perez2024stability}, and neural networks \cite{xu2020active, lee2019data, fukami2020assessment} to predict aerodynamic coefficients, vortex shedding frequencies, and flow field evolution for stationary and, to a lesser extent, rotating cylinders \cite{bairagi2024artificial}. Lee {\it et al.} \cite{lee2019data} demonstrated that physics-informed networks give more accurate and physically consistent predictions while adversarial training improves the capture of unsteady flow features like vortex shedding. Using ANN, thermophysical properties of magnetohydrodynamics flow with radiation in an arc-shaped enclosure with a rotating cylinder were predicted by Bairagi {\it et al.} \cite{bairagi2024artificial} with a mean square error of 0.00069, resulting in a 99\% accuracy rate. Fukami {\it et al.} \cite{fukami2020assessment} adopted four machine learning algorithms of multilayer perceptron, random forest, support vector regression, and extreme learning machine for stationary cylinder wake dynamics. Furthermore, the authors considered use of convolutional neural network in the context of super-resolution analysis, and found it to be a suitable tool to handle big data. However, supervised machine learning models were not found to be suitable for extrapolation problems \cite{yu2019flowfield, sengupta2025benchmark}. While these approaches have demonstrated promising capabilities in capturing nonlinear dependencies and providing rapid predictions once trained. However, the majority of existing supervised machine learning-based studies are confined to incompressible flows and relatively low-dimensional parameter spaces. Applications to rotating cylinders have primarily focused on lift and drag prediction under steady or weakly unsteady conditions \cite{hasegawa2020machine, asghar2024predicting}, with limited attention to highly nonlinear regimes involving bifurcations and multi-frequency interactions.

To the best of the authors’ knowledge, machine learning frameworks have not been systematically applied to high-fidelity compressible flow past rotating cylinders at high rotation rates, where the flow exhibits strong nonlinearities, intermittent transitions, and multi-scale interactions. The challenges associated with such flows, such as, including sensitivity to input parameters, non-monotonic response behavior, and high computational cost of data generation, make them particularly well-suited for data-driven modeling, yet underexplored in current literature. Thus, the present study combines high-fidelity compressible simulations with supervised machine learning techniques to investigate flow past a rotating cylinder across a wide range of $Re_{\infty}$ in regimes characterized by strong nonlinear dynamics. The work aims not only to characterize the underlying flow physics, including bifurcation behavior and spectral features, but also to develop robust data-driven surrogate and generative models capable of predicting key flow quantities with high accuracy. In doing so, this study seeks to bridge the gap between traditional high-fidelity simulations and modern machine learning approaches for complex compressible flow systems.

\FloatBarrier\section{Problem formulation for compressible flow past a rotating cylinder at a fixed high rotation rate}
\label{sec:2}
\begin{figure}[htbp]
\centering
\includegraphics[width=\textwidth]{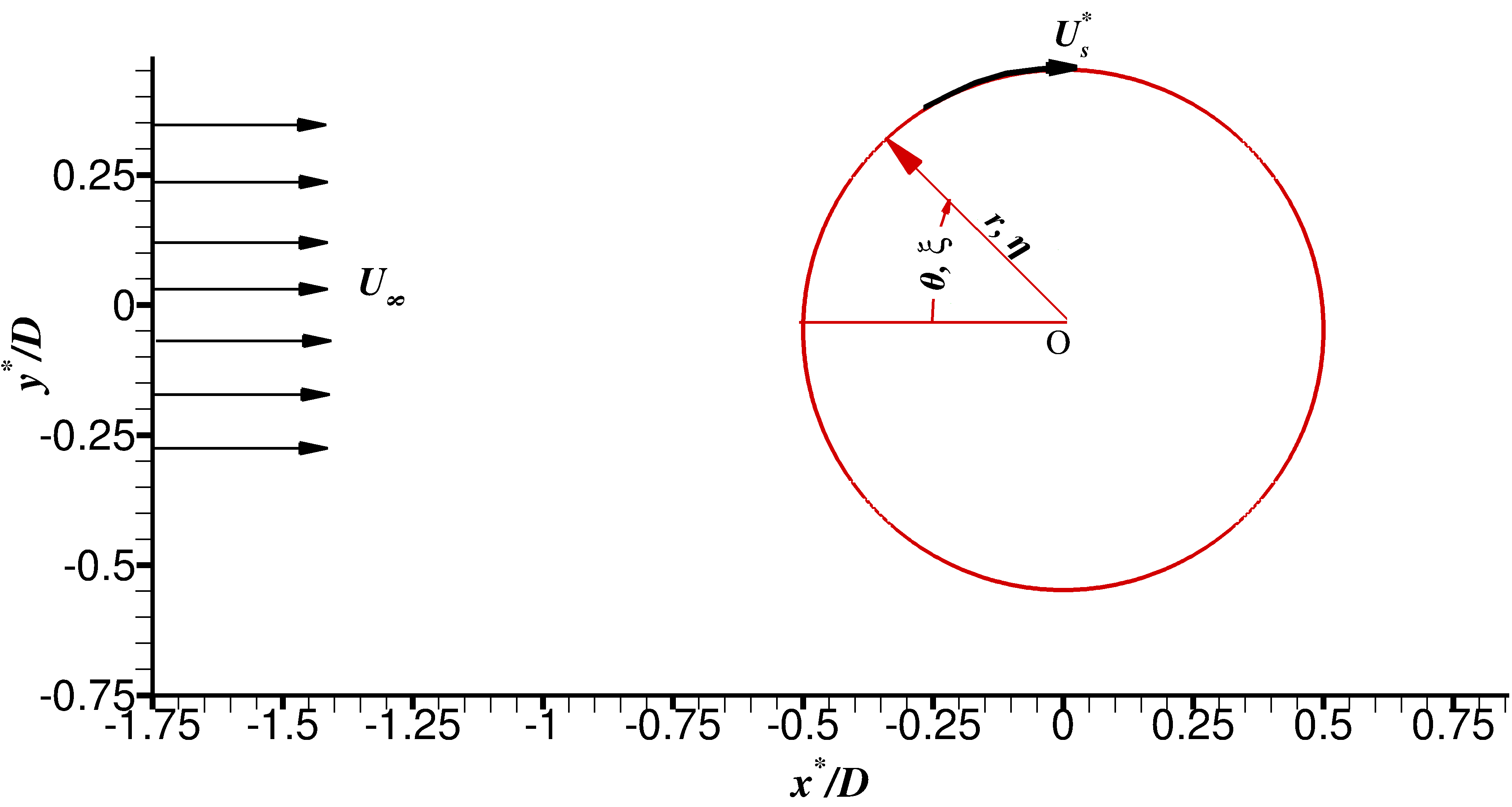}
\caption{Schematic of uniform flow past a rotating cylinder with surface speed, $U^*_s$.}
\label{fig1}
\end{figure}

The schematic for two-dimensional flow past a rotating cylinder in free stream is given in Fig. \ref{fig1}. The flow field is generated by an infinitely long cylinder of diameter, $D$ which is rotating about its axis with angular velocity, $\Omega^*$.  The uniform flow having velocity, $U_{\infty}$ is translating from left to right and impinging the rotating cylinder at right angles to its axis. The azimuthal angle ($\theta$) is marked with respect to the most upstream point of the cylinder on the windward side. Instead of quantifying the rotation in terms of a rotation rate, here, it is indicated by the tangential surface speed, $U^*_s$, evaluated as $U^*_s = \Omega^* D/2$. On the surface of the cylinder we prescribe no-slip and adiabatic boundary conditions. The far-field is taken at 30$D$ where characteristic based \citep{pulliam1986implicit} boundary condition is used.

We solve the two-dimensional compressible Navier-Stokes equations, formulated by established notations in \citep{hoffmann2000computational}.

\begin{eqnarray}
\frac{\partial {\hat{Q}}}{\partial t} + \frac{\partial {\hat{E}}}{\partial x} + \frac{\partial {\hat{F}}}{\partial y} = \frac{\partial {\hat{E_{v}}}}{\partial x} + \frac{\partial {\hat{F_{v}}}}{\partial y}
\label{eq2.1} 
\end{eqnarray}

\noindent where the conservative variables are given as, $\hat{Q} = [ \rho \;\; \rho u \;\; \rho v \;\; e_t]^T$. The convective flux vectors are similarly given as,

\begin{equation}
\hat{E} = [ \rho u \;\; \rho u^2 +p \;\; \rho uv \;\; (\rho e_t + p) u]^T 
\label{eq2.2} 
\end{equation}
\begin{equation}
\hat{F} = [ \rho v \;\; \rho uv \;\; \rho v^2 +p \;\; (\rho e_t + p) v]^T
\label{eq2.3} 
\end{equation}

\noindent and the viscous flux vectors are given as,

\begin{equation}
\hat{E_{v}}= [ 0\;\; \tau _{xx} \;\; \tau _{xy} \;\; (u \tau _{xx} +v \tau _{xy} -q _{x})]^T
\label{eq2.4} 
\end{equation}
\begin{equation}
\hat{F_{v}}= [0 \;\; \tau _{yx} \;\; \tau _{yy} \;\; (u \tau _{yx} +v \tau _{yy} -q _{y})]^T
\label{eq2.5} 
\end{equation}

In Eqs. \eqref{eq2.1} to \eqref{eq2.5}, $\rho$, $u$, $v$, $e_{t}$, and $p$ denote dimensionless values of density, velocity components, total specific energy, and pressure, respectively. These physical variables are normalized with respect to the free-stream density ($\rho_{\infty}$), free-stream velocity ($U_{\infty}$), free-stream temperature ($T_{\infty}$), free-stream dynamic viscosity ($\mu_{\infty}$), length scale ($D$), and the time scale, ($D/U_{\infty}$). The dimensionless parameters, namely the Prandtl number ($Pr$), free-stream Reynolds number ($Re_{\infty}$), and free-stream Mach number, ($M_{\infty}$), are defined as follows:

\begin{equation*}
Pr = \frac{\mu_{\infty} C_{p}}{\kappa}	;\ Re_{\infty} = \frac{\rho_{\infty} U_{\infty} D}{\mu_{\infty}}	;\ M_{\infty} = \frac{U_{\infty}}{\sqrt{\gamma R^* T_{\infty}}}
\end{equation*}    

\noindent where $\gamma=1.4$ represents the ratio of specific heat capacity at constant pressure ($C_p$) to constant volume ($C_v$). The pressure scale ($p_{\infty}$) is evaluated as $p_{\infty} = \rho_{\infty} R^* T_{\infty}$, where $R^*$ is the dimensional gas constant, non-dimensionalized as $R = R^*T_{\infty}/U^2_{\infty}$. The equation of state for an ideal gas, $p = \rho R T$ is used to relate the state variables and $e_t$ is defined as $e_t = C_v T + \frac{1}{2} (u^2 + v^2)$. Heat conduction terms are given by

\begin{equation*}
q_{x}= - \frac{\mu}{Pr Re_{\infty} (\gamma -1) M_{\infty}^2} \frac{\partial {T}}{\partial x}; \; \;\\
q_{y}= - \frac{\mu}{Pr Re_{\infty} (\gamma -1) M_{\infty}^2} \frac{\partial {T}}{\partial y}
\end{equation*}

\noindent The components of the Newtonian viscous stress tensors, $\tau_{xx},\tau_{xy},\tau_{yx},\tau_{yy}$, are defined as 

\begin{equation*}
\tau_{xx} = \frac{1}{Re_{\infty}}\Big[ 2\mu \frac{\partial u}{\partial x} + \lambda \nabla \cdot \vec{V} \Big]; \; \\
\tau_{yy} = \frac{1}{Re_{\infty}}\Big[2\mu\frac{\partial v}{\partial y}+ \lambda \nabla \cdot \vec{V} \Big]; \; \\
\tau_{xy} = \tau_{yx} = \frac{\mu}{Re_{\infty}}\Big[\frac{\partial u}{\partial y} + \frac{\partial v}{\partial x} \Big]
\end{equation*}

\noindent Here, Stokes' hypothesis ($\lambda = -\frac{2}{3} \mu $) is used in calculating stress tensor. To calculate viscosity as a function of temperature, Sutherland's law is applied. The equations from the Cartesian space ($x$, $y$) are transformed to body-fitted computational grid ($\xi, \eta$) using the following relations: $\xi = \xi (x, y)$ and $\eta = \eta (x, y)$. The transformed plane equations in strong conservation form are given as, 

\begin{eqnarray}
\frac{\partial {{\bf Q}}}{\partial t} + \frac{\partial {\bf E}}{\partial \xi} + \frac{\partial {\bf F}}{\partial \eta} = \frac{\partial {\bf E_{v}}}{\partial \xi} + \frac{\partial {\bf F_{v}}}{\partial \eta}
\label{eq6} 
\end{eqnarray}   
\noindent with the state variables and flux vectors, given as 

\begin{equation*}
{\bf Q} = \hat{Q}/J; \; \; \; \\
{\bf E} = (\xi _{x} \hat{E} +\xi _{y} \hat{F})/J; \; \; \; \\
{\bf F} = (\eta _{x} \hat{E} +\eta _{y} \hat{F})/J; \; \; \;\\
\end{equation*}
\begin{equation}
 {\bf E_{v}} = (\xi _{x} \hat{E_{v}} +\xi _{y} \hat{F_{v}})/J; \; \\
{\bf F_{v}} = (\eta _{x} \hat{E_{v}} +\eta _{y} \hat{F_{v}})/J   
\end{equation}

\noindent and $J$ is the Jacobian of the grid transformation given by $
J=\frac{1}{x_{\xi} y_{\eta} -x_{\eta} y_{\xi}}$. The grid metrics, $\xi_{x}$, $\xi_{y}$, $\eta_{x}$, and $\eta_{y}$, are computed during the creation of an O-type grid using a hyperbolic grid generation technique in Pointwise. These are expressed as follows: $\xi_{x} = J y_{\eta}$; $\xi_{y} = -J x_{\eta}$; $\eta_{x} = -J y_{\xi}$; $\eta_{y} = J x_{\xi}$ and are used in calculating partial derivatives of the Cartesian grid in the transformed plane. For the computational domain shown in Fig. \ref{fig1}, an $O$-type grid consisting of 401 equidistant points in $\xi$-direction and 450 points in $\eta$-direction is adopted. In the $\eta$-direction, grid points are clustered near the wall using the following tangent hyperbolic stretching function: $$r(\eta) = 0.5 + r_{max} \biggl[ 1-\frac{tanh\beta \biggl (\frac{N_{\eta}-\eta}{N_{\eta}-1}\biggr)}{tanh \beta}\biggr]$$ where $N_{\eta}$ is number of points in $\eta$-direction, $r_{max}$ is determined from far-field boundary (set to 30$D$ in present simulations), and $\beta$ is stretching parameter which controls wall spacing, $\Delta r_{wall}$. Previous work reported on rotating cylinder \citep{sundaram2021multiscale} performed a mesh independence study using three wall resolutions, $\Delta r_{wall}$ = 0.0005$D$, 0.001$D$ and 0.002$D$. For the first two wall spacings, the coefficient of lift ($C_l$) was found to be identical. Hence, for the test cases reported here, we use the finest resolution, i.e. $\Delta r_{wall} = 0.0005D$ for which $\beta = 3$.

\subsection{Numerical methodology}

The governing equations in Eq. \eqref{eq6} are solved by first, discretizing the convective flux terms in space. In the present work, this is achieved using the optimized upwind compact scheme OUCS3, which provides near-spectral accuracy across a broad range of wavenumbers compared with conventional discretization approaches \citep{sagaut2023global}. The viscous flux derivatives are computed using a non-uniform explicit central difference formulation. To suppress high-wavenumber numerical oscillations, a fourth-order artificial diffusion term proposed by \cite{jameson2017origins} is introduced with a coefficient of 0.015. Time integration is carried out using the optimized three-stage Runge–Kutta scheme (OCRK3) \citep{sagaut2023global} with a non-dimensional time step of $\Delta t = 1.25 \times 10^{-5}$. The resulting space–time discretization has been examined through global spectral analysis, demonstrating that the numerical schemes preserves the dispersion relation linking spatial and temporal scales \cite{zhou2007unification} in the flow. The in-house finite-difference-based compressible Navier–Stokes solver is parallelized using a novel high-accuracy non-overlapping parallel (NOHAP) algorithm. This has been adopted to completely eliminate errors at sub-domain boundaries between successive processors without the need for large number of overlapping points. This parallel algorithm has been benchmarked for a variety of incompressible and compressible problems \cite{sengupta2024thermal, sundaram2023non, joshi2024highly, joshi2025comparing}, showing identical results as sequential computing. 

The objective of the present study is to conduct a bifurcation analysis of the flow past a rotating cylinder operating at a high rotation rate, where the surface velocity is prescribed as $U^*_s = 10 U_{\infty}$. The free-stream Reynolds number, $Re_{\infty}$, is varied between 1000 and 6000 in increments of 50. In the vicinity of the bifurcation point ($Re_{\infty}$ > 5300), additional simulations are performed with finer $Re_{\infty}$-number spacing, resulting in a total of 144 two-dimensional compressible flow simulations. The free-stream Mach number is maintained at $M_{\infty}$ = 0.1, ensuring that the flow remains within the incompressible regime. The Prandtl number is fixed at $Pr = 0.71$, representative of air, while the reference temperature is $T_{\infty} = 288.15$K, corresponding to a density scale of $\rho_{\infty} = 1.2256kg/m^3$.

The numerical framework employed here has previously been validated against the experimental measurements of \cite{tokumaru1993lift} in \citep{sundaram2021multiscale}, where simulations at $Re_{\infty} = 3800$ and dimensionless rotation rates ranging from 0.5 to 10 $U_{\infty}$ showed good agreement with the experimentally measured mean lift coefficient. Furthermore, the normalized transverse velocity for the current flow parameters were also compared with experiments of Tokumaru and Dimotakis \cite{tokumaru1993lift} in \cite{sengupta2025bifurcation}, demonstrating a good agreement. Each simulation in the present dataset is executed using 40 cores for approximately 240 computational hours, continuing until limit-cycle oscillations are established in the lift and drag coefficients. The complete set of 144 simulations therefore requires approximately 1.4 million core hours. This extensive computational effort is undertaken to generate a high-fidelity benchmark dataset for the flow past a rapidly rotating cylinder across a wide range of $Re_{\infty}$. 

\FloatBarrier
\section{Results and Discussion}

\label{sec:3}
This section investigates the vorticity dynamics of flow past a rotating cylinder across different $Re_{\infty}$. The dominant temporal scales are identified using Fast Fourier Transform (FFT) analysis of the lift and drag signals. The study then explores supervised learning approaches to characterize the bifurcation behavior at high rotation rates, comparing polynomial and Bayesian regression models for predicting drag coefficient, lift coefficient, and the onset time of instability. An artificial neural network (ANN) model is subsequently developed to estimate critical flow parameters, providing an efficient surrogate to expensive high-fidelity simulations. Finally, the ability of the ANN models to generalize beyond the training dataset is demonstrated.

\subsection{Role of Reynolds number in vorticity dynamics for flow past a rapidly rotating cylinder}

Rapid rotation strongly affects the boundary-layer development and separation behaviour in flow past a rotating cylinder. On the side where the surface motion is aligned with the free stream, the boundary layer experiences a favorable pressure gradient and tends to remain attached for a longer arc length. On the opposite side, where the surface motion opposes the free stream, the boundary layer decelerates more rapidly and separates earlier. This asymmetry shifts the wake and produces a deflected shear layer that contributes to the net circulation around the cylinder \cite{sengupta2003temporal}. 

\begin{figure}[!ht]
\centering
\includegraphics[width=\textwidth]{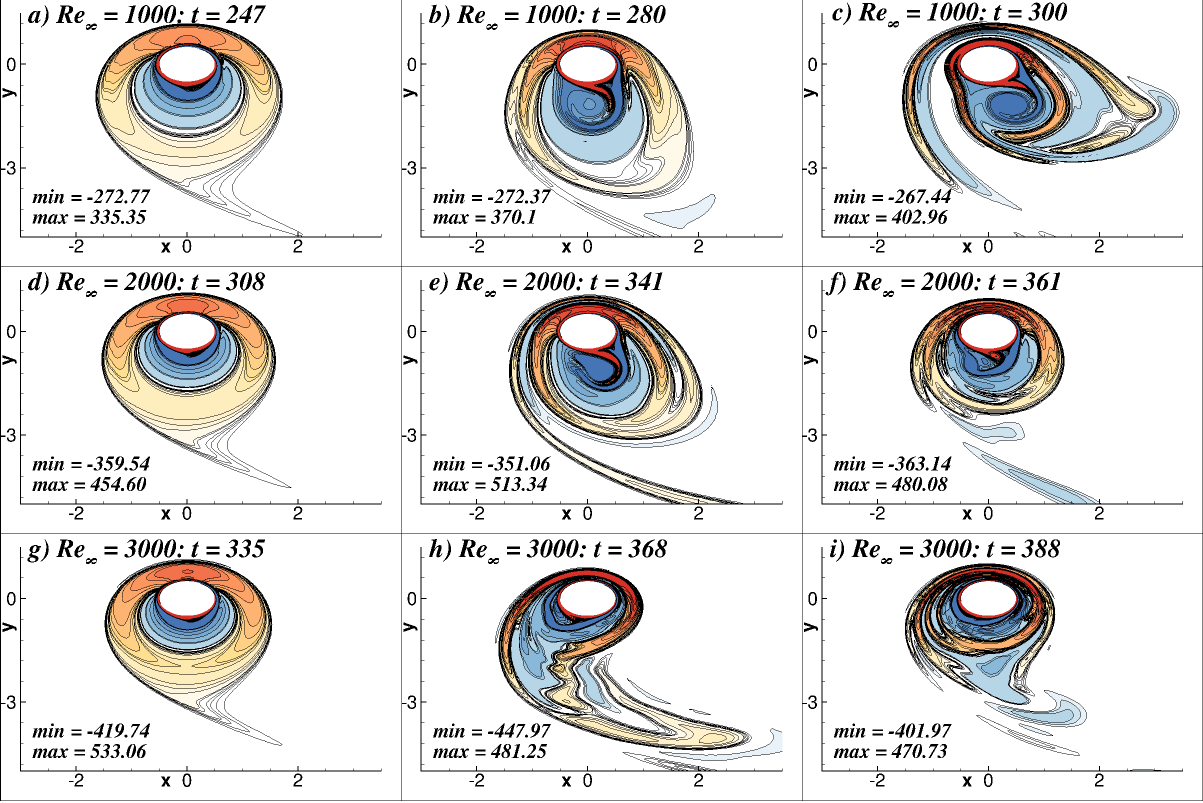}
\caption{Spanwise vorticity contours for flow past a rotating cylinder with $U^*_s = 10U_{\infty}$ and (a)-(c)$Re_{\infty} = 1000$, (d)-(f)$Re_{\infty} = 2000$, (g)-(i) $Re_{\infty} = 3000$. The onset of the instability is shown in frames (a),(d),(g).}
\label{fig2}
\end{figure}

The time variation of spanwise vorticity contours are compared for $Re_{\infty} = 1000$ to 3000 in Fig. \ref{fig2} starting from the onset time of instability in the leftmost frames. For a cylinder rotating at a high non-dimensional rotation rate ($U_s^* = 10$), the flow field is strongly influenced by asymmetric boundary-layer development around the cylinder. The rotation accelerates the flow on the upper surface and decelerates it on the lower surface relative to the free stream. As a result, the upper side typically remains largely attached due to favorable pressure gradients, while the lower side experiences stronger adverse pressure gradients that promote boundary-layer thickening and separation \cite{pal2026direct}. The streamline topology therefore exhibits characteristic saddle points and separation points in the vicinity of the lower surface and in the near wake. At sufficiently high rotation rates, the wake becomes asymmetric: the shear layer originating from the lower side of the cylinder rolls up into vortical structures, while the upper shear layer remains relatively stable and attached \cite{munir2021flow}. This leads to the formation of concentrated regions of positive and negative vorticity in the near wake. When the flow becomes unsteady, these shear layers undergo vortex roll-up and periodic shedding, which manifests as oscillations in the aerodynamic forces.

The spanwise vorticity contours in Figs. \ref{fig2}(a)–(c), \ref{fig2}(d)–(f), and \ref{fig2}(g)–(i) correspond to $Re_{\infty} = 1000$, 2000 and 3000, respectively. The onset of the temporal instability for each $Re_{\infty}$ is depicted in Figs. \ref{fig2}(a), \ref{fig2}(d) and \ref{fig2}(g). At this time-instant, the vorticity distributions for all three $Re_{\infty}$ appear nearly identical. This similarity indicates that the initial instability mechanism is governed primarily by the global shear-layer structure induced by rotation, which is largely independent of $Re_{\infty}$ within this range \cite{chew1995numerical}. At the onset stage, the flow is dominated by the same large-scale separation pattern and wake topology, leading to comparable vorticity distributions. As time progresses, however, clear differences emerge. The cases with higher $Re_{\infty}$ exhibit stronger unsteadiness, particularly in the lower portion of the cylinder where the adverse pressure gradient and shear-layer development are most pronounced, as shown in Figs. \ref{fig2}(b), \ref{fig2}(e) and \ref{fig2}(h). At higher $Re_{\infty}$, viscous diffusion is weaker, allowing the separated shear layer to remain thinner and more unstable. This enhances the amplification of perturbations in the shear layer, leading to earlier roll-up and stronger vortex formation. Consequently, the flow at higher $Re_{\infty}$ generates more intense positive vorticity near the lower surface and in the downstream wake. The increased inertial effects promote stronger roll-up of the separated shear layer, producing a more energetic and unsteady wake \cite{sengupta2020effects_a}. In contrast, at lower $Re_{\infty}$ the higher viscous dissipation smooths the vorticity gradients, delaying shear-layer instability and resulting in comparatively weaker vortex activity.

\begin{figure}[!ht]
\centering
\includegraphics[width=\textwidth]{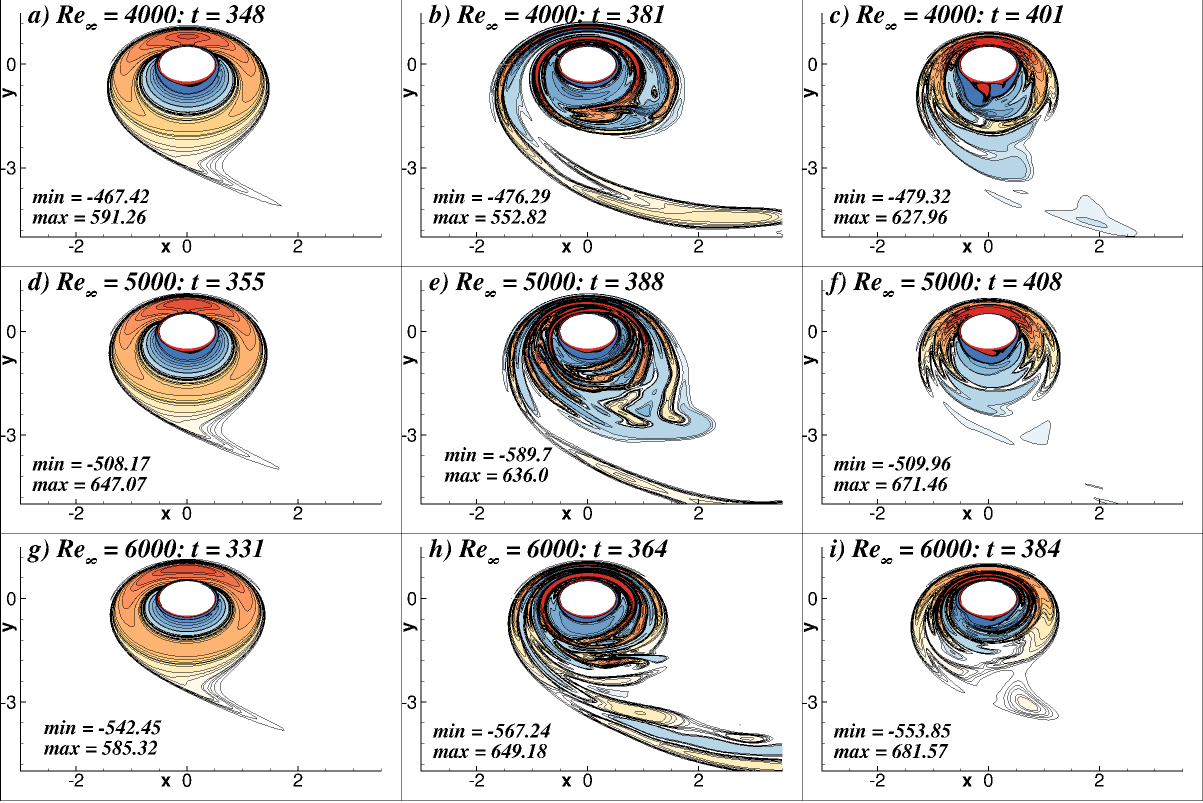}
\caption{Spanwise vorticity contours for flow past a rotating cylinder with $U^*_s = 10U_{\infty}$ and (a)-(c)$Re_{\infty} = 4000$, (d)-(f)$Re_{\infty} = 5000$, (g)-(i) $Re_{\infty} = 6000$. The onset of the instability is shown in frames (a),(d),(g).}
\label{fig3}
\end{figure}

The time variation of spanwise vorticity contours are compared in Figs. \ref{fig3}(a)–(c), \ref{fig3}(d)–(f), and \ref{fig3}(g)–(i) corresponding to $Re_{\infty} = 4000$, 5000 and 6000, respectively. The onset of the temporal instability for each $Re_{\infty}$ is depicted in Figs. \ref{fig3}(a), \ref{fig3}(d) and \ref{fig3}(g). At the onset of temporal instability, the vorticity fields retain the same qualitative structure observed in Fig. \ref{fig2}. The flow exhibits an asymmetric wake due to the high rotation rate, with the upper shear layer remaining relatively attached and the lower shear layer separating and forming the primary region of vorticity production. The streamline topology shows similar separation patterns and shear-layer formation around the cylinder. For the higher $Re_{\infty}$ shown here, the separated shear layer becomes noticeably thinner and more unstable at later time instants. Compared with the earlier cases ($Re_{\infty} \le 3000$) in Fig. \ref{fig2}, the roll-up of the shear layer occurs more rapidly and produces stronger vortical structures around the lower portion of the cylinder. This behavior is consistent with the reduction in viscous diffusion at higher $Re_{\infty}$, which allows steeper velocity gradients and stronger vorticity concentrations to develop. 

In Figs. \ref{fig3}(b), \ref{fig3}(e) and \ref{fig3}(h), the shear layer wraps around the cylinder and begins forming large coherent vortices. Relative to the lower $Re_{\infty}$ shown in Fig. \ref{fig2}, the vortex cores appear more tightly wound and more concentrated, indicating stronger vorticity production. The wake of the cylinder also shows increased curvature and distortion, suggesting stronger interaction between the rotating boundary layer and the separated wake \cite{sengupta2025bifurcation}. At later times in Figs. \ref{fig3}(c), \ref{fig3}(f) and \ref{fig3}(i), the wake exhibits greater fragmentation and smaller-scale vortical structures, particularly for $Re_{\infty} = 5000$ and 6000. The increased inertial dominance at these $Re_{\infty}$ promotes stronger shear-layer roll-up, leading to a more chaotic wake \cite{richter2012effects, sengupta2025compressible}. In contrast, the lower $Re_{\infty}$ in Fig. \ref{fig2} displayed comparatively smoother vortical structures and slower wake evolution due to stronger viscous damping.

\subsection{Unsteady force distribution for various $Re_{\infty}$-flows past a rapidly rotating cylinder}

The lift generated by a rapidly rotating cylinder in a uniform crossflow arises primarily from the Magnus-Robins effect \cite{kundu2012fluid}, which results from the interaction between the cylinder’s rotation and the incoming flow. When the cylinder rotates, the tangential velocity of the surface adds to the free-stream velocity on one side of the cylinder and subtracts from it on the opposite side. This produces an asymmetric velocity distribution around the cylinder and consequently an asymmetric pressure field. The pressure difference across the cylinder generates a net transverse force, which appears as lift.

\begin{figure}[!ht]
\centering
\includegraphics[width=\textwidth]{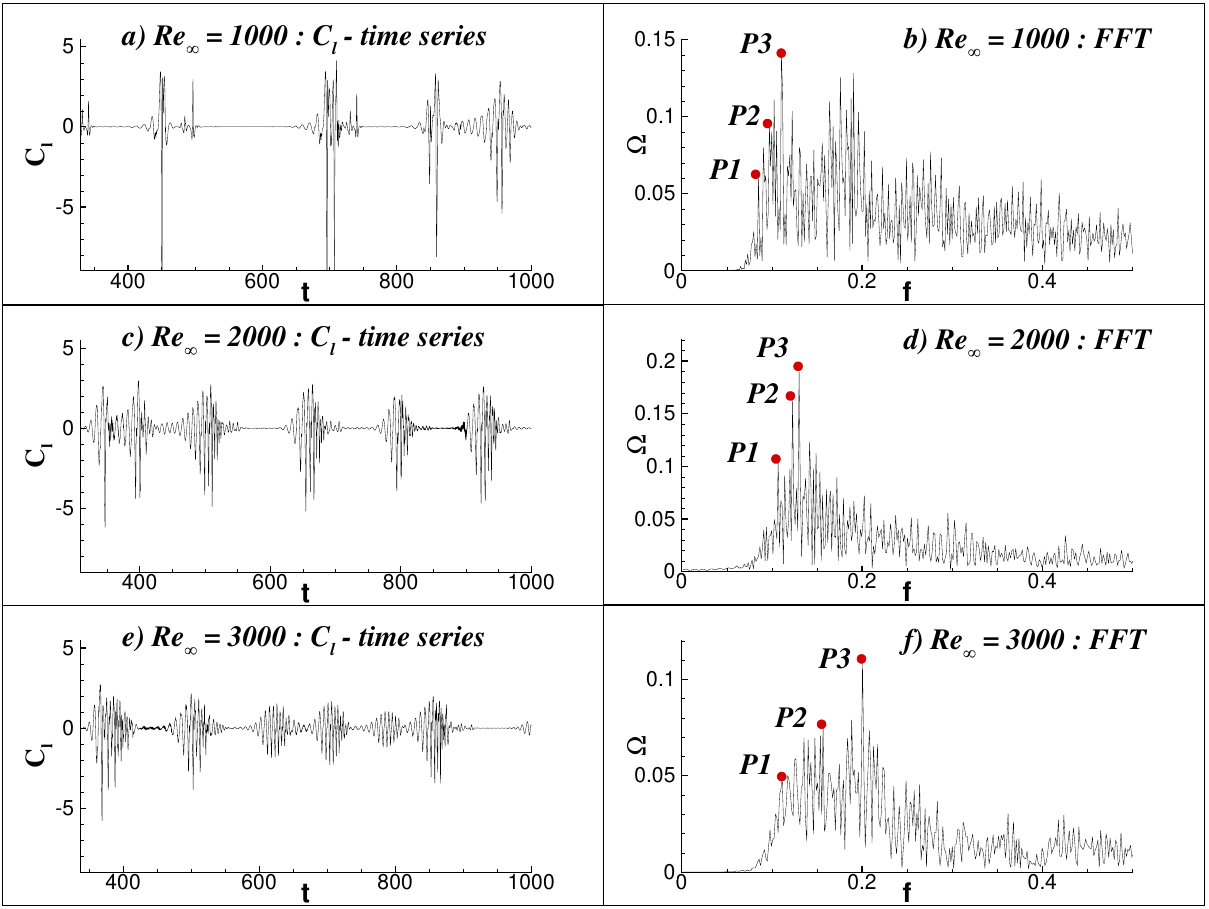}
\caption{Time variation of lift coefficient for (a) $Re_{\infty} = 1000$, (c) $Re_{\infty} = 2000$, (e) $Re_{\infty} = 3000$ and their corresponding spectra are shown in frames (b), (d), and (f).}
\label{fig4}
\end{figure}

The time histories of the lift coefficient ($C_l$) and their corresponding FFT spectra in Fig. \ref{fig4} reveal the dominant temporal scales associated with the wake dynamics at $Re_{\infty} = 1000$, 2000 and 3000. For all three Reynolds numbers, the lift signal exhibits periodic oscillations once the flow reaches the limit-cycle state, indicating the presence of coherent vortex shedding. The FFT spectra confirm this behavior by showing a dominant peak P$_1$, which represents the primary shedding frequency, along with secondary peaks P$_2$ and P$_3$ that correspond to higher harmonics of the fundamental mode. The observed spectral features can be interpreted in light of the vortical structures presented in Fig. \ref{fig2}. The onset of instability is associated with the separation and roll-up of the lower shear layer caused by the high rotation rate of the cylinder. This process generates a coherent vortex structure that periodically interacts with the wake, leading to oscillations in the aerodynamic forces \cite{suman2022novel}. The dominant peak P$_1$ in the FFT therefore corresponds to the fundamental frequency associated with this primary vortex-shedding mechanism. The presence of the additional peaks P$_2$ and P$_3$ can be attributed to nonlinear interactions within the separated shear layer. As the vortices convect downstream and interact with the wake, harmonic components of the fundamental shedding frequency are generated, which appear as secondary peaks in the frequency spectrum. This behavior is consistent with the vorticity contours in Fig. \ref{fig2}, where the shear layer undergoes progressive roll-up and deformation as time evolves.

A comparison across Reynolds numbers reveals that although the fundamental frequency P$_1$ persists for all cases, the relative amplitudes of the secondary peaks become more pronounced with increasing $Re_{\infty}$. This trend reflects the stronger shear-layer instability and increased wake unsteadiness observed in Fig. \ref{fig2} for higher Reynolds numbers. At larger $Re_{\infty}$, reduced viscous damping allows sharper vorticity gradients and stronger vortex interactions to develop in the lower wake region, which enhances nonlinear mode coupling and leads to more energetic harmonic content in the lift spectrum.

\begin{figure}[!ht]
\centering
\includegraphics[width=\textwidth]{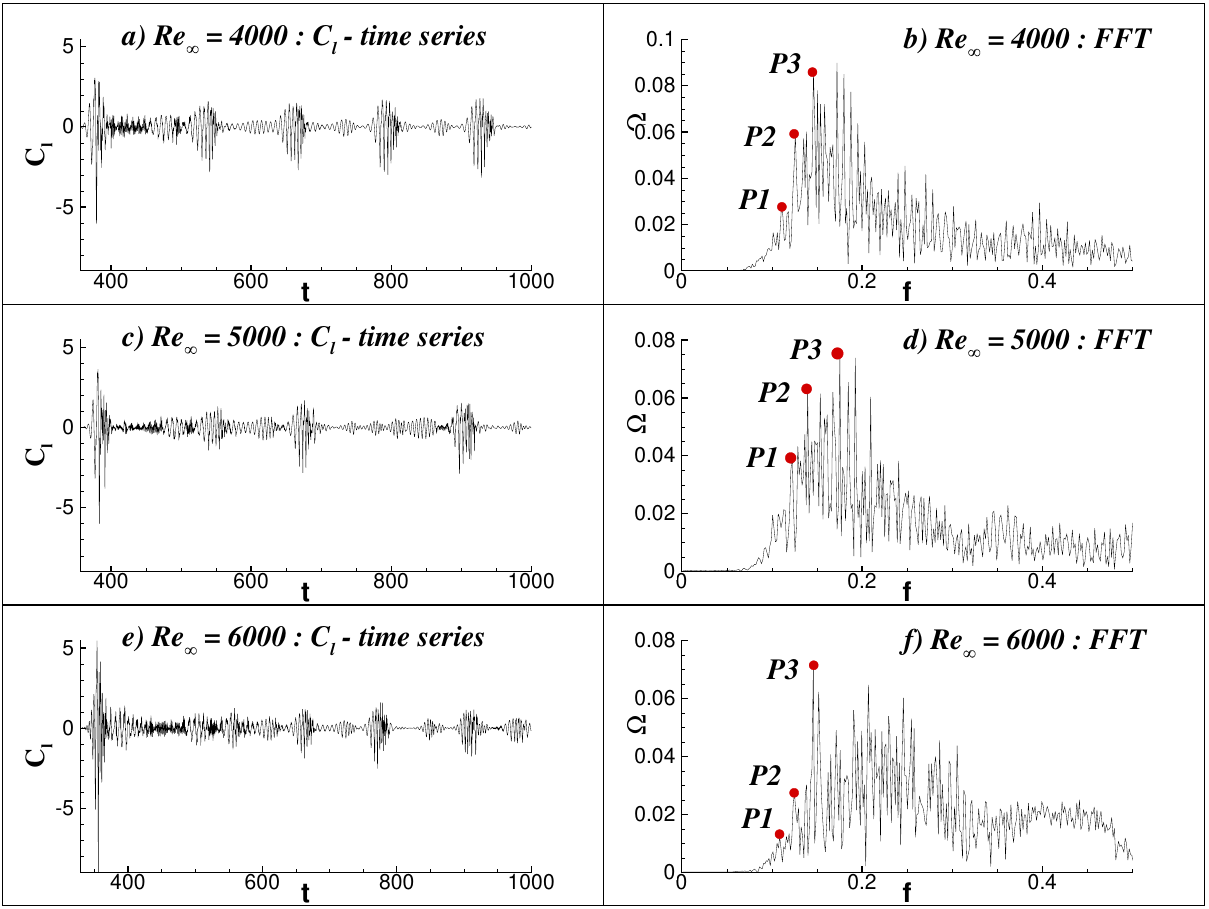}
\caption{Time variation of lift coefficient for (a) $Re_{\infty} = 4000$, (c) $Re_{\infty} = 5000$, (e) $Re_{\infty} = 6000$ and their corresponding spectra are shown in frames (b), (d), and (f).}
\label{fig5}
\end{figure}

The time-series of $C_l$ and corresponding FFT spectra for $Re_{\infty} = 4000$, 5000, and 6000 are depicted in Fig. \ref{fig5} which show several differences when compared with the lower-$Re_{\infty}$-cases of Fig. \ref{fig4}. The time-series of lift distribution at higher $Re_{\infty}$ exhibits more irregular oscillations compared with the relatively smoother periodic signals observed at lower $Re_{\infty}$. While a clear limit-cycle behavior is still present, the waveform shows subtle distortions, indicating stronger nonlinear interactions in the wake. This behavior is consistent with the vorticity field of Fig. \ref{fig3}, where higher $Re_{\infty}$ produced thinner and more unstable shear layers and more energetic vortex roll-up. The FFT spectra reveal that although the dominant peak $P_1$ is present for all $Re_{\infty}$, its relative amplitude decreases for the higher $Re_{\infty}$. The secondary peaks $P_2$ and $P_3$ become comparatively more pronounced, especially $P_3$ for post-bifurcation \cite{sengupta2025bifurcation} case with $Re_{\infty} = 6000$ in Fig. \ref{fig5}(d). The lift signal contains stronger harmonic content at higher $Re_{\infty}$, resulting from nonlinear interactions among vortical structures in the wake, which become more significant as viscous damping decreases. Another notable difference is the broader spectral distribution seen for higher $Re_{\infty}$, compared with Fig. \ref{fig4}, where the spectral energy is concentrated mainly at the fundamental frequency and its harmonics. This reflects the increased complexity of the wake dynamics, where vortex merging, and interactions generate additional temporal scales in the flow for higher $Re_{\infty}$.

At $Re_{\infty}$ = 6000, the flow lies beyond the bifurcation threshold ($Re_{\infty}$ > 5650) reported in \cite{sengupta2025bifurcation}. This fundamentally alters the temporal dynamics of the wake. In the pre-bifurcation regime, the lift signal is primarily governed by a single dominant vortex-shedding mode, represented by the peak $P_1$ in the FFT spectrum. The secondary peaks $P_2$ and $P_3$ arise mainly from harmonic generation caused by nonlinear interactions within the separated shear layer. However, post-bifurcation wake dynamics transition to a new oscillatory state characterized by mode competition and energy redistribution among frequencies \cite{pralits2010instability}. In this regime, the primary instability associated with the original shedding frequency loses dominance, and a higher-frequency mode becomes energetically favorable. This shift is reflected in the FFT spectrum for $Re_{\infty}$ = 6000, where the peak $P_3$ exhibits the largest amplitude.

Physically, this behavior can be attributed to the increased instability of the separated shear layer along the lower surface of the rotating cylinder. At higher $Re_{\infty}$, the shear layer becomes thinner and more susceptible to Kelvin–Helmholtz type instabilities. These instabilities generate smaller-scale vortical structures and introduce additional temporal scales into the wake \cite{munir2021flow}. As bifurcation occurs, the flow reorganizes such that the dominant lift fluctuations are no longer governed solely by the large-scale vortex shedding but are instead influenced by higher-frequency shear-layer dynamics. Consequently, the energy in the lift signal shifts toward higher frequencies, causing $P_3$ to dominate the spectrum. This change indicates that the flow has transitioned from a single-frequency limit-cycle oscillation to a multi-mode oscillatory state \cite{barkley1996three}, where nonlinear interactions between wake vortices and shear-layer instabilities redistribute the spectral energy.

Table \ref{tab1} summarizes the three most energetic frequencies identified from the FFT of $C_l$ in Figs. \ref{fig4} and \ref{fig5}. For each $Re_{\infty}$, the frequencies corresponding to the three largest spectral peaks are denoted as $P_1$, $P_2$ and $P_3$, while $\Omega$ represents the amplitude of the lift signal at those frequencies.

\begin{table}[H]
\begin{center}
\def~{\hphantom{0}}
\caption{Dominant frequencies and their amplitudes, marked as P1, P2, P3 in Figs. \ref{fig4} and \ref{fig5}}

\begin{tabular}{lccccccc}
 \hline
$Re_{\infty}$ & P1($f$) & P1($\Omega$) & P2($f$) & P2($\Omega$) & P3($f$) & P3($\Omega$)  \\ [3pt] 
 \hline 
 1000 & 0.085333 & 0.061313 & 0.097333 & 0.094369 & 0.110666 & 0.139604 \\ 
 \hline
 2000 & 0.107093 & 0.104940 & 0.123010 & 0.164021 & 0.130246 & 0.191612 \\
 \hline
 3000 & 0.108792 & 0.048307 & 0.156862 & 0.075814 & 0.200603 & 0.109376 \\ 
 \hline
 4000 & 0.110769 & 0.028033 & 0.1261538 & 0.059012 & 0.1461538 & 0.0856537 \\
 \hline
 5000 & 0.111613 & 0.039488 & 0.1397515 & 0.062551 & 0.175465 & 0.075116 \\
 \hline
 6000 & 0.122670 & 0.012176 & 0.1251862 & 0.0272304 & 0.1460506 & 0.0708161 \\
 \hline
\end{tabular}
\label{tab1}
\end{center}
\end{table}

A clear trend in Table \ref{tab1} is that the dominant frequency $P_1$(f) increases slightly with $Re_{\infty}$, rising from approximately 0.085 at $Re_{\infty}$ = 1000 to about 0.123 at $Re_{\infty}$ = 6000. This gradual increase reflects the strengthening of the vortex-shedding process as inertial effects become more important with increasing $Re_{\infty}$. The higher inertia leads to thinner separated shear layers and faster vortex formation, which in turn increases the characteristic shedding frequency.

For the lower $Re_{\infty} \le 3000$, the fundamental mode $P_1$ typically dominates the spectrum, while $P_2$ and $P_3$ appear as higher harmonics generated by nonlinear interactions in the wake. The amplitudes of these secondary peaks remain smaller than the primary peak, indicating that the lift oscillations are primarily governed by a single vortex-shedding mode with relatively weak harmonic contributions. As the $Re_{\infty}$ increases in range (4000-5000), the spectral distribution begins to change. Although $P_1$ is present, the amplitudes of $P_2$ and $P_3$ become more comparable, suggesting stronger nonlinear interactions within the separated shear layer. This behavior is consistent with the increasingly unstable wake structures observed in the vorticity contours in Fig. \ref{fig3}. A more pronounced shift occurs at $Re_{\infty} = 6000$, which lies in the post-bifurcation regime. In this case, the amplitude associated with $P_3$ exceeds that of $P_1$, indicating that the higher-frequency mode becomes energetically dominant. This transition reflects a redistribution of energy among the wake instability modes following the bifurcation. 

The drag generated in the flow past a rapidly rotating cylinder results from the combined effects of pressure forces associated with wake formation and viscous shear stresses along cylinder surface. For a stationary cylinder in crossflow, drag is dominated by pressure (form) drag, which arises due to separation of the boundary layer and formation of a low-pressure wake behind the cylinder \cite{white1994fluid}. When the cylinder rotates rapidly, there is asymmetric separation which produces a deflected wake and modifies the pressure distribution around the cylinder. At high rotation rates, the wake is often shifted and compressed toward one side of the cylinder which can reduce pressure drag. Viscous drag, on the other hand, arises from shear stress acting on the cylinder surface due to the tangential velocity gradient within the boundary layer. Rapid rotation increases magnitude of surface velocity and therefore enhances wall shear stresses, particularly on the side where surface moves opposite to the free stream. As a result, viscous contribution to the total drag becomes more significant than in stationary-cylinder case, although pressure drag remains the dominant component at moderate to high $Re_{\infty}$.

\begin{figure}[!ht]
\centering
\includegraphics[width=\textwidth]{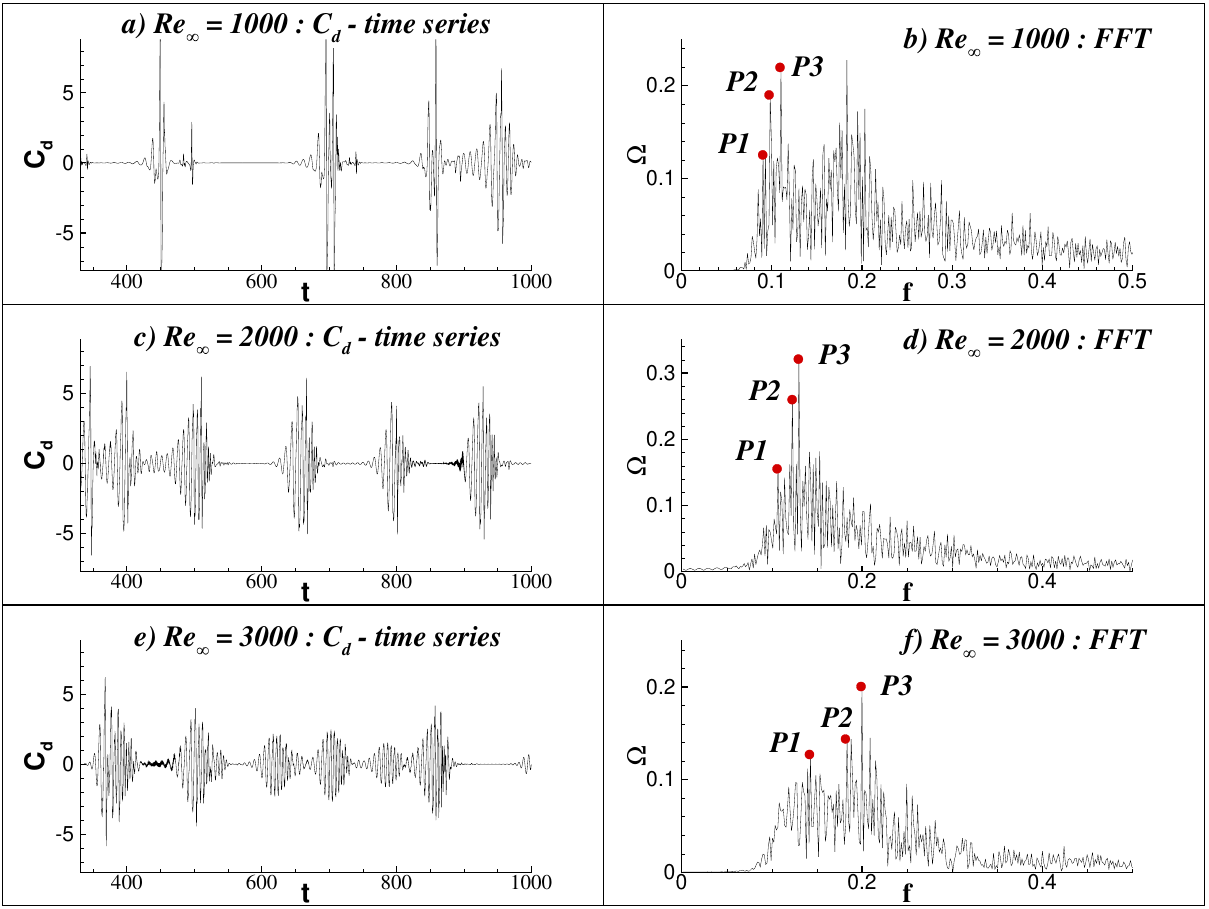}
\caption{Time variation of drag coefficient for (a) $Re_{\infty} = 1000$, (c) $Re_{\infty} = 2000$, (e) $Re_{\infty} = 3000$ and their corresponding spectra are shown in frames (b), (d), and (f).}
\label{fig6}
\end{figure}

The time histories of the drag coefficient, $C_d$ and their corresponding FFT spectra for $Re_{\infty} = 1000$, 2000, and 3000 in Fig. \ref{fig6} reveal how the unsteady drag response evolves with $Re_{\infty}$. The drag signals exhibit periodic oscillations once the flow reaches the limit-cycle state, indicating that the drag fluctuations are closely linked to the periodic formation and convection of vortical structures in the wake. The onset of instability is usually characterized by the beginning of this limit-cycle oscillation phase. At $Re_{\infty} = 1000$ shown in Fig. \ref{fig6}(a), the time-series of $C_d$ is relatively smooth and exhibits nearly periodic oscillations with a well-defined frequency. The corresponding FFT spectrum in Fig. \ref{fig6}(b) shows a clear dominant peak $P_1$, which represents the primary frequency associated with the vortex formation process in the wake. The additional peaks, $P_2$ and $P_3$ appear at higher frequencies and correspond to harmonic components generated by nonlinear interactions in the separated shear layer. As $Re_{\infty}$ increases to 2000 in Fig. \ref{fig6}(c), the overall amplitude of the drag oscillations becomes larger and the waveform begins to show distortions compared with the $Re_{\infty} = 1000$ case. The FFT spectrum in Fig. \ref{fig6}(d) still displays a dominant peak, $P_1$, but the secondary peaks, $P_2$ and $P_3$ become more energetic. This indicates that the wake dynamics involve stronger nonlinear interactions between vortices, resulting in additional harmonic content in the drag signal. For the $C_d$ time-series at $Re_{\infty} = 3000$ in Fig. \ref{fig6}(e), the drag fluctuations exhibit more pronounced variability and the waveform deviates further from a sinusoidal pattern. The corresponding FFT spectrum in Fig. \ref{fig6}(f) shows that the energy distribution among the peaks becomes more spread out, with $P_2$ and $P_3$ contributing more significantly to the total spectral energy. This behavior reflects the increasing instability of the separated shear layers at higher $Re_{\infty}$ as viscous diffusion becomes weaker, and thereby shear layers are thinner and more susceptible to instability.

\begin{figure}[!ht]
\centering
\includegraphics[width=\textwidth]{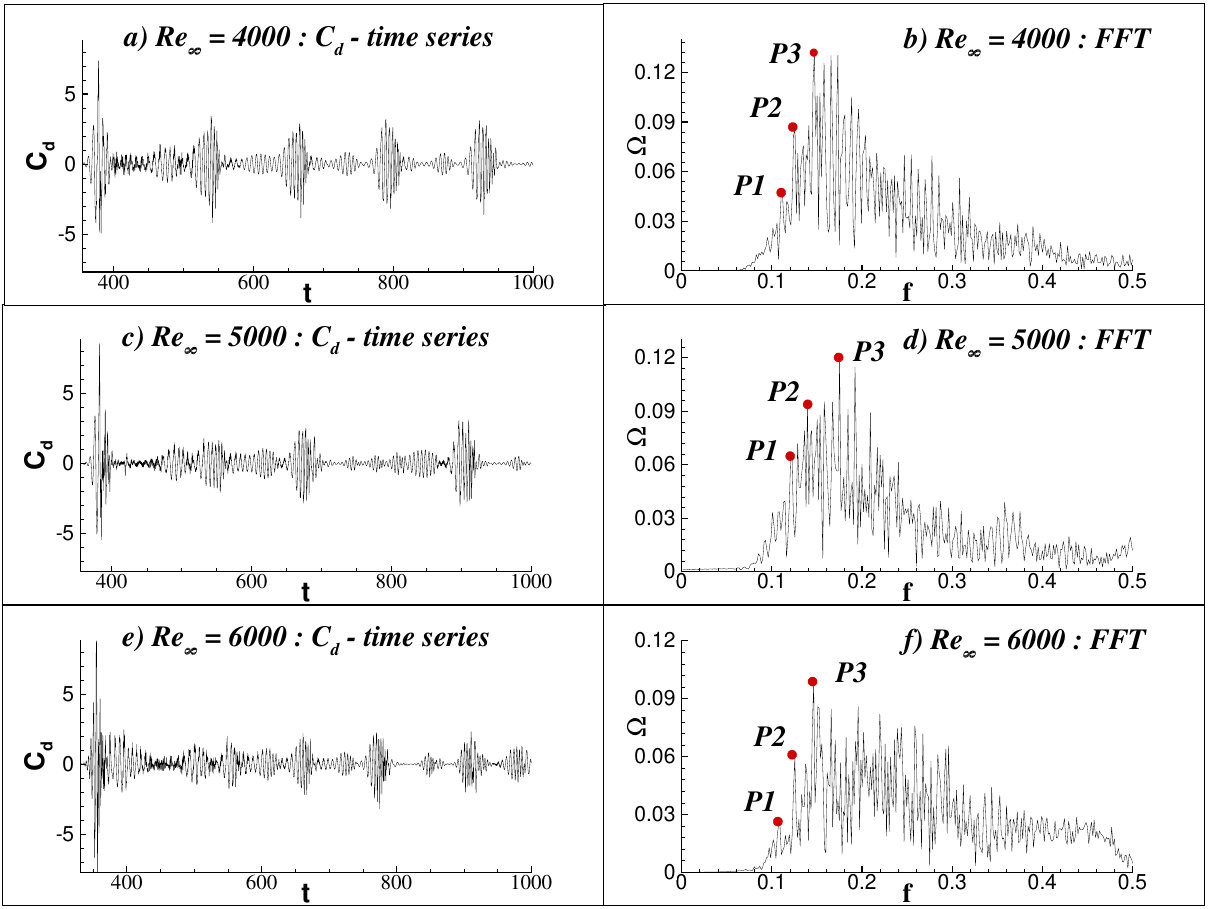}
\caption{Time variation of drag coefficient for (a) $Re_{\infty} = 4000$, (c) $Re_{\infty} = 5000$, (e) $Re_{\infty} = 6000$ and their corresponding spectra are shown in frames (b), (d), and (f).}
\label{fig7}
\end{figure}

The time histories of $C_d$ for $Re_{\infty} = 4000$, 5000 and 6000 together with their FFT spectra illustrated in Fig. \ref{fig7} show the evolution of the drag dynamics in the higher-$Re_{\infty}$ regime. For $Re_{\infty} = 4000$ depicted in Fig. \ref{fig7}(a), the time series shows near-periodic oscillations with moderate amplitude. There is an overall reduction in $C_d$ for these higher $Re_{\infty}$ as although stronger shear-layer instabilities are promoted, the strong rotational forcing stabilizes part of the boundary layer and shifts the wake asymmetrically. As a result, periodic variations in the streamwise pressure difference across the cylinder become smaller, leading to reduced drag fluctuations. The corresponding FFT spectrum in Fig. \ref{fig7}(b) displays the typical peak $P_1$ indicative of the primary frequency associated with the vortex formation process in the wake, but having lower relative dominance. The peaks $P_2$ and $P_3$ appear at higher frequencies and correspond to harmonic components generated by nonlinear interactions within the separated shear layers. At higher $Re_{\infty}$, the wake instability's energy is redistributed among multiple frequencies, as indicated by the FFT spectra. Instead of a single dominant mode producing large oscillations, the drag signal contains several interacting modes with comparable amplitudes. This spectral redistribution spreads the fluctuation energy over a broader range of frequencies, which reduces the amplitude of individual oscillations in the time signal. As $Re_{\infty}$ increases to 5000 in Fig. \ref{fig7}(c), the drag oscillations become more distorted with pronounced aperiodicity, while the energy distribution in the FFT of Fig. \ref{fig7}(d) begins to broaden over a larger frequency range. While the fundamental peak $P_1$ remains identifiable, the amplitudes of the secondary peaks, $P_2$ and $P_3$ are of almost comparable magnitude to $P_1$. This indicates stronger nonlinear coupling between the wake vortices and the separated shear layers. The reduced viscous diffusion at higher $Re_{\infty}$ allows sharper velocity gradients and more energetic vortex interactions, which in turn introduce additional temporal scales into the drag signal. A more pronounced change occurs for $Re_{\infty} = 6000$, which lies in the post-bifurcation regime. The time-series of $C_d$ in Fig. \ref{fig7}(e) displays noticeable modulation in the oscillations, reflecting the presence of multiple interacting instability modes. Consistent with this behaviour, the FFT spectrum in Fig. \ref{fig7}(f) shows a redistribution of spectral energy among the three peaks, with the higher-frequency components $P_3$ and $P_2$ becoming more significant. This trend contrasts with the lower $Re_{\infty}$ depicted in Fig. \ref{fig6}, where the spectral energy is primarily concentrated at the fundamental frequency, $P_1$.

Table \ref{tab2} summarizes the three most energetic frequencies extracted from the FFT of $C_d$ for all $Re_{\infty}$. The terms $P_1$, $P_2$, and $P_3$ denote the three dominant peaks identified in the spectra shown in Figs. \ref{fig6} and \ref{fig7}, while $\Omega$ represents the corresponding spectral amplitude of the drag signal.

\begin{table}[ht]
\begin{center}
\def~{\hphantom{0}}
\caption{Dominant frequencies and their amplitudes, marked as P1, P2, P3 in Figs. \ref{fig6} and \ref{fig7}}
\begin{tabular}{lccccccc}
 \hline
$Re_{\infty}$ & P1($f$) & P1($\Omega$) & P2($f$) & P2($\Omega$) & P3($f$) & P3($\Omega$)  \\ [3pt] 
 \hline 
 1000 & 0.0929612 & 0.126777 & 0.0982733 & 0.186975 & 0.1102256 & 0.216485 \\ 
 \hline
 2000 & 0.1069364 & 0.155420 & 0.1228323 & 0.260650 & 0.1300578 & 0.318439 \\
 \hline
 3000 & 0.1428571 & 0.124854 & 0.1834580 & 0.144039 & 0.1999990 & 0.197753 \\ 
 \hline
 4000 & 0.1104436 & 0.045746 & 0.1257660 & 0.085343 & 0.1472392 & 0.1309634 \\
 \hline
 5000 & 0.1209312 & 0.064209 & 0.1395348 & 0.092911 & 0.1751937 & 0.119128 \\
 \hline
 6000 & 0.1076235 & 0.024942 & 0.1255604 & 0.059098 & 0.146487 & 0.097337 \\
 \hline
\end{tabular}
\label{tab2}
\end{center}
\end{table}

For $1000 \le Re_{\infty} \le 3000$, the amplitudes of the spectral peaks are relatively large. In particular, the higher-frequency components, $P_2$ and $P_3$ exhibit amplitudes that are comparable to or larger than the fundamental peak, $P_1$. This indicates that the drag signal contains significant harmonic content arising from nonlinear interactions between vortical structures in the wake. As $Re_{\infty}$ increases to 2000, the amplitudes of all three peaks increase, reflecting stronger vortex formation and enhanced fluctuations in the wake pressure field. At $Re_{\infty} = 3000$, the spectral energy becomes more distributed among the three modes, suggesting stronger nonlinear coupling between the primary shedding frequency and its higher harmonics.

A notable change occurs for $4000 \le Re_{\infty} \le 6000$, with an overall decrease in amplitudes of all spectral peaks. This can be attributed to the strong rotational forcing of the cylinder, which modifies the wake topology and suppresses large-scale streamwise pressure fluctuations. At higher $Re_{\infty}$, boundary layer on the accelerating side of the cylinder remains attached for a longer distance, producing a narrower and more deflected wake and thereby reducing magnitude of drag oscillations. Despite reduction in amplitude, dominant frequencies remain within a relatively narrow band across the entire $Re_{\infty}$ range. This indicates that the characteristic temporal scales of the wake dynamics are preserved, even though the strength of the drag fluctuations changes. For $Re_{\infty} = 6000$, which lies in the post-bifurcation regime, the spectral energy becomes more evenly distributed among the peaks, reflecting the presence of multiple interacting instability modes \cite{zhou2024hydrodynamic} in the wake.

\subsection{Polynomial regression applied to bifurcation analysis} 

\begin{figure}[!ht]
\centering
\includegraphics[width=.7\textwidth]{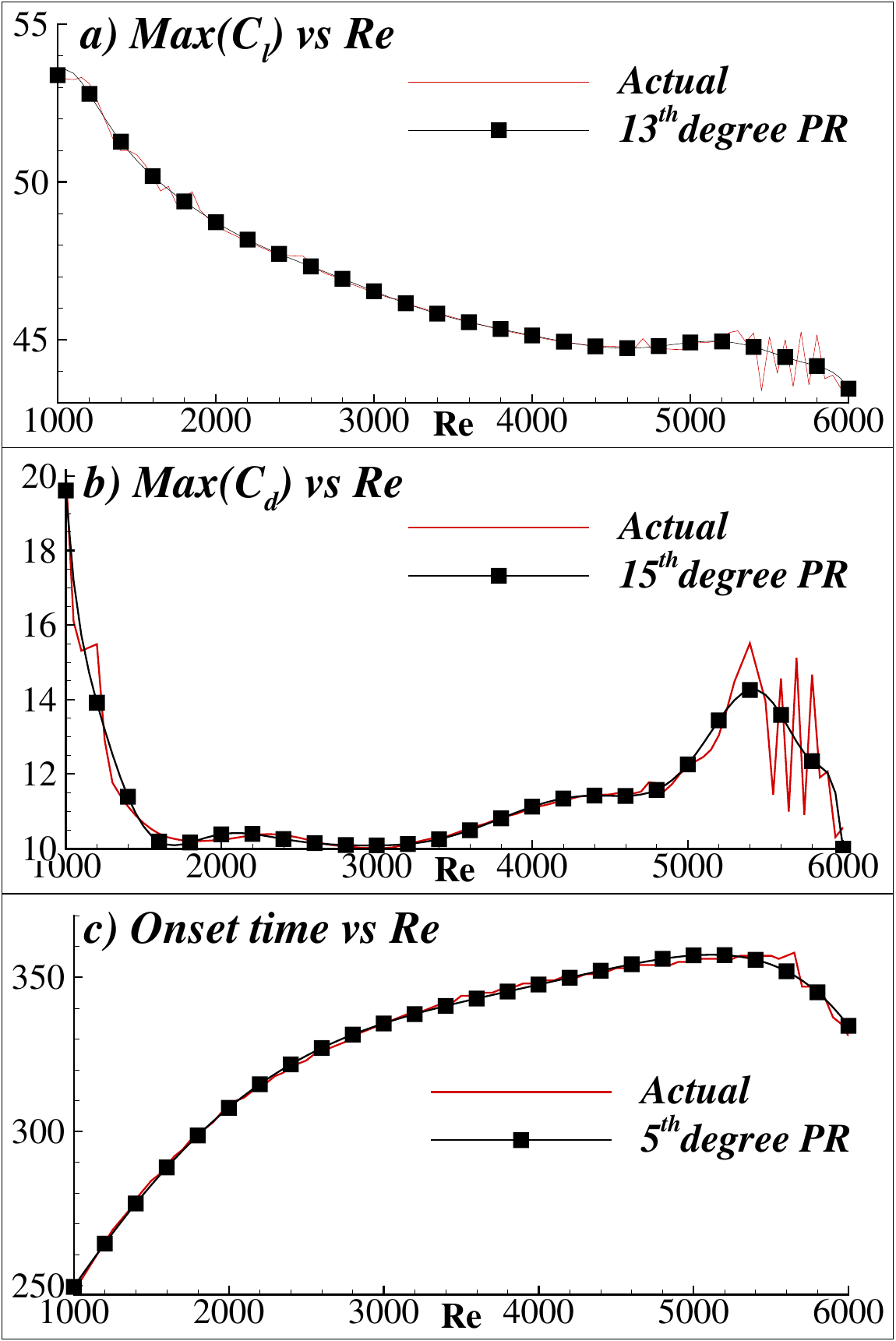}
\caption{Variation of (a) maximum $C_l$, (b) maximum $C_d$, and (c) onset time of instability, with $Re_{\infty}$, along with polynomial curve fits of different degrees.}
\label{fig8}
\end{figure}

To characterize the dependence of the critical flow parameters on $Re_{\infty}$, polynomial regression is employed as a simple supervised learning approach. Polynomial regression extends linear regression by representing the relationship between an input variable and a response variable as a polynomial function, thereby enabling the model to capture nonlinear trends in the data, as previously reported in \cite{gond2025data}. In the present study, this method is applied to approximate the variation of three key quantities: the maximum lift coefficient, maximum drag coefficient, and the onset time of instability — as functions of $Re_{\infty}$, as depicted in Fig. \ref{fig8}. These quantities exhibit nonlinear variation with $Re_{\infty}$ due to changes in wake dynamics and shear-layer instability. Polynomial regression therefore provides a convenient data-driven framework to identify smooth trends in the simulation data and to construct predictive relationships between $Re_{\infty}$ and the resulting aerodynamic response.

Figure \ref{fig8} shows a smooth variation of (a) maximum $C_l$, (b) maximum $C_d$, and (c) onset time of instability across most of the $Re_{\infty}$ range, reflecting the gradual strengthening of wake dynamics as inertial effects increase. However, a pronounced change in behavior occurs near $Re_{\infty} \approx 5650$, where sharp fluctuations appear in both the maximum lift and drag distributions, and the onset-time curve departs from its earlier parabolic trend and bifurcates, as reported in \cite{sengupta2025bifurcation}. This behavior indicates the presence of a dynamical bifurcation \cite{theofilis2003advances} in the wake structures of the cylinder. For $Re_{\infty} < 5650$, the flow evolves toward a relatively stable limit-cycle oscillation governed by a single dominant instability mode associated with vortex formation in the separated shear layer. In this regime, the aerodynamic forces vary smoothly with $Re_{\infty}$ because the underlying wake topology changes gradually.

As $Re_{\infty}$ approaches the critical value, the separated shear layer becomes increasingly thin and susceptible to secondary instabilities. Small perturbations in the shear layer amplify rapidly, leading to stronger vortex roll-up and enhanced vortex–vortex interactions \cite{sundaram2021multiscale}. Once the critical $Re_{\infty}$ is exceeded, the wake transitions to a post-bifurcation state characterized by multiple interacting instability modes. In this regime, the force coefficients become highly sensitive to small variations in $Re_{\infty}$, producing the sharp fluctuations observed in the maximum lift and drag curves. The bifurcation in the onset-time curve can be interpreted in the same framework. Prior to the transition, the onset of instability follows a smooth trend because the growth rate of the dominant instability varies gradually with $Re_{\infty}$. Beyond the critical point, however, the wake dynamics are governed by competing instability mechanisms. As a result, the time required for perturbations to amplify and produce observable force oscillations no longer follows a single monotonic trend, leading to the branching behavior observed in Fig. \ref{fig8}(c). The growth rate of unstable modes increases sharply and the flow can no longer suppress disturbances with rotation alone. Global instability modes dominate the dynamics and the flow transitions from linearly stable to fully nonlinear and unsteady.

Overall, the sharp variations in maximum $C_l$, maximum $C_d$, and the bifurcation in the onset time around $Re_{\infty} \approx 5650$ signify a transition in the wake dynamics from a single-mode oscillatory state to a more complex multi-mode regime, consistent with the spectral behavior observed in the lift and drag spectra of Figs. \ref{fig4} to \ref{fig7}.

Table 3 summarizes the error metrics used to determine the optimal polynomial degree for regression models predicting the maximum $C_l$, maximum $C_d$, and the onset time of instability as functions of $Re_{\infty}$. Three statistical indicators are used to evaluate model performance: the mean absolute error (MAE), the mean squared error (MSE), and the prediction accuracy. The MAE measures the average absolute deviation between the predicted and simulated values, while the MSE penalizes larger deviations more strongly by squaring the error. Lower values of both MAE and MSE indicate a better fit. The reported accuracy quantifies how closely the regression predictions match the simulation data.

\begin{longtable}{|c|c|c|c|c|}
\caption{Mean absolute error (MAE), mean square error (MSE), and accuracy achieved with various degrees of polynomial for onset time, maximum $C_l$ and maximum $C_d$, as a function of $Re_{\infty}$.} \\
\hline
\textbf{Case} & \textbf{Degree} & \textbf{MAE} & \textbf{MSE} & \textbf{Accuracy (\%)} \\ 
\hline
\endfirsthead

\hline
\textbf{Case} & \textbf{Degree} & \textbf{MAE} & \textbf{MSE} & \textbf{Accuracy (\%)} \\ 
\hline
\endhead

& 1 & 0.8827 & 1.1373 & 82.51 \\ \cline{2-5}
& 2 & 0.3553 & 0.2314 & 96.44 \\ \cline{2-5}
& 3 & 0.2479 & 0.1110 & 98.29 \\ \cline{2-5}
& 4 & 0.2480 & 0.1110 & 98.29 \\ \cline{2-5} 
& 5 & 0.1717 & 0.0872 & 98.65 \\ \cline{2-5}
& 6 & 0.1615 & 0.0859 & 98.67 \\ \cline{2-5}
Max $C_l$ vs $Re_{\infty}$ & 7 & 0.1628 & 0.0825 & 98.73 \\ \cline{2-5}
& 8 & 0.1659 & 0.0818 & 98.74 \\ \cline{2-5}
& 9 & 0.1487 & 0.0794 & 98.77 \\ \cline{2-5}
& 10 & 0.1537 & 0.0778 & 98.80 \\ \cline{2-5}
& 11 & 0.1537 & 0.0777 & 98.80 \\ \cline{2-5}
& 12 & 0.1525 & 0.0758 & 98.83 \\ \cline{2-5}
& {\bf 13} & {\bf 0.1532} & {\bf 0.0757} & {\bf 98.84} \\ \hline

& 1 & 1.1787 & 3.0503 & 3.13 \\ \cline{2-5}
& 2 & 0.9275 & 1.9103 & 39.33 \\ \cline{2-5}
& 3 & 0.7641 & 1.0967 & 65.16 \\ \cline{2-5}
& 4 & 0.7395 & 1.0588 & 66.37 \\ \cline{2-5}
& 5 & 0.5405 & 0.5387 & 82.89 \\ \cline{2-5}
& 6 & 0.5370 & 0.5327 & 83.08 \\ \cline{2-5}
& 7 & 0.3346 & 0.4258 & 86.47 \\ \cline{2-5}
& 8 & 0.3494 & 0.4203 & 86.65 \\ \cline{2-5}
& 9 & 0.3220 & 0.3990 & 87.32 \\ \cline{2-5}
Max $C_d$ vs $Re_{\infty}$ & 10 & 0.2932 & 0.3930 & 87.51 \\ \cline{2-5}
& 11 & 0.3048 & 0.3922 & 87.54 \\ \cline{2-5}
& 12 & 0.3124 & 0.3889 & 87.64 \\ \cline{2-5}
& 13 & 0.3146 & 0.3659 & 88.37 \\ \cline{2-5}
& 14 & 0.3081 & 0.3657 & 88.38 \\ \cline{2-5}
& {\bf 15} & {\bf 0.2950} & {\bf 0.3622} & {\bf 88.49} \\ \hline

& 1 & 11.2691 & 204.6 & 74.67 \\ \cline{2-5} 
& 2 & 2.9782 & 13.976 & 98.29 \\ \cline{2-5} 
& 3 & 2.5503 & 12.214 & 98.50 \\ \cline{2-5} 
Onset time vs $Re_{\infty}$ & 4 & 1.2952 & 3.7407 & 99.54 \\ \cline{2-5}
& {\bf 5} & {\bf 0.6462} & {\bf 2.3052} & {\bf 99.87} \\ \cline{2-5}
& 6 & 1.1085 & 1.0188 & 99.71 \\ \cline{2-5}
& 7 & 0.5058 & 0.7917 & 99.03 \\ \hline
\end{longtable}

For the maximum $C_l$, the regression accuracy improves rapidly when moving from a linear model (degree 1) to higher-order polynomials. The MAE decreases from 0.8827 for the linear fit to approximately 0.15 for higher-degree polynomials, while the MSE decreases by more than an order of magnitude. Beyond approximately degree 6–7, the improvement becomes marginal, indicating that the regression has already captured the essential nonlinear dependence of maximum $C_l$ on $Re_{\infty}$. The highest accuracy of 98.84\% with the lowest MSE is achieved for the 13th-degree polynomial, which is therefore selected as the best-performing model.

For the maximum $C_d$, the relationship with $Re_{\infty}$ is more complex and requires a slightly higher polynomial order to achieve an accurate fit. Both MAE and MSE decrease gradually with increasing polynomial degree, and the prediction accuracy improves from only 3.13\% for the linear model to 88.49\% for the 15th-degree polynomial. This higher degree reflects the stronger fluctuations in drag observed near the bifurcation region, which require a more flexible regression model to capture the nonlinear variation.

For the onset time of instability, the regression converges much more quickly. The MAE drops dramatically from 11.27 for the linear model to below 1 for moderate polynomial degrees, and the accuracy rapidly exceeds 99\%. The 5th-degree polynomial provides the best balance between error reduction and model complexity, achieving 99.87\% accuracy with a significantly lower MAE compared to lower-order fits. Higher-degree polynomials do not yield consistent improvements and may introduce unnecessary complexity.

The optimal polynomial degree varies depending on the physical quantity being modeled. The lift and drag maxima exhibit stronger nonlinear dependence on $Re_{\infty}$, requiring higher-order polynomials, whereas the onset time follows a smoother trend that can be captured effectively with a lower-degree polynomial.

Table \ref{tab4} summarizes the final polynomial regression models obtained for the three key response variables: maximum $C_l$, maximum $C_d$, and the onset time of instability, as functions of $Re_{\infty}$ the free-stream Reynolds number. In each case, the polynomial degree was selected based on the error analysis presented in Table 3, ensuring the best balance between predictive accuracy and model complexity.

\begin{table}[h!]
\centering
\caption{Polynomial regression applied to onset time, maximum $C_l$ and maximum $C_d$ as a function of $Re_{\infty}$.}
\renewcommand{\arraystretch}{1.2} 
\begin{tabular}{|c|c|}
\hline
\textbf{Cases} & \textbf{Polynomial Expression} \\
\hline
Max $C_l$ vs $Re_{\infty}$ & 
\scriptsize
$\begin{aligned}
y(x) = &3.639\times10^{-43}x^{13} - 1.882\times10^{-38}x^{12} + 4.352\times10^{-34}x^{11} - 5.964\times10^{-30}x^{10} \\
       & + 5.406\times10^{-26}x^{9} - 3.422\times10^{-22}x^{8} + 1.557\times10^{-18}x^{7} - 5.148\times10^{-15}x^{6} \\
       &  + 1.237\times10^{-11}x^{5} - 2.134\times10^{-8}x^{4} + 2.564\times10^{-5}x^{3} - 2.032\times10^{-2}x^{2} \\
       &   + 9.503x - 1.921\times10^{3}
\end{aligned}$  \\
\hline
Max $C_d$ vs $Re_{\infty}$ & 
\scriptsize
$\begin{aligned}
y(x) = &-1.521\times10^{-48}x^{15} + 7.974\times10^{-44}x^{14} - 1.913\times10^{-39}x^{13} + 2.706\times10^{-35}x^{12} \\
       &- 2.784\times10^{-31}x^{11} + 1.997\times10^{-27}x^{10} - 1.066\times10^{-23}x^{9} + 4.301\times10^{-20}x^{8} \\
       &- 1.323\times10^{-16}x^{7} + 3.094\times10^{-13}x^{6} - 5.455\times10^{-10}x^{5} + 7.107\times10^{-7}x^{4} \\
       &- 6.615\times10^{-4}x^{3} + 4.148\times10^{-1}x^{2} - 1.568\times10^{2}x + 2.684\times10^{4}
\end{aligned}$  \\
\hline
Onset time vs $Re_{\infty}$ & 
\scriptsize
$\begin{aligned}
y(x) = &-3.057\times10^{-16}x^{5} + 4.4085\times10^{-12}x^{4} - 2.1561\times10^{-8}x^{3} + 3.1308\times10^{-5}x^{2} \\
       &+ 5.8456\times10^{-2}x + 1.7724\times10^{2}
\end{aligned}$  \\
\hline
\end{tabular}
\label{tab4}
\end{table}  

For the maximum $C_l$, the optimal model corresponds to a 13th-degree polynomial, as depicted in Fig. \ref{fig8}(a). This relatively high-order polynomial reflects the strongly nonlinear dependence of maximum $C_l$ on $Re_{\infty}$, particularly near the bifurcation region where abrupt changes in wake dynamics lead to sharp variations in the lift response. The polynomial coefficients progressively decrease in magnitude for higher-order terms, indicating that the dominant contribution arises from the lower-order terms, while the higher-order terms primarily provide local corrections that enable the regression model to accurately capture subtle variations in the simulation data.

For the maximum $C_d$, a 15th-degree polynomial shown in Fig. \ref{fig8}(b) provides the best fit. The higher polynomial order compared to the lift model suggests that the drag response exhibits even stronger nonlinear behavior with $Re_{\infty}$. This is consistent with the physical observations that drag is influenced by both pressure and viscous effects, as well as by complex wake interactions that intensify near the bifurcation point. The additional polynomial degrees allow the regression model to represent the fluctuating trends across $Re_{\infty}$ while maintaining acceptable prediction accuracy.

In contrast, the onset time of instability follows a smoother variation with $Re_{\infty}$ and can be accurately represented using a 5th-degree polynomial, as shown in Fig. \ref{fig8}(c). The relatively low polynomial order indicates that the dependence of the instability onset time on $Re_{\infty}$ is more systematic and less oscillatory compared to the aerodynamic force coefficients. The dominant contribution arises from the lower-order terms of the polynomial, which capture the overall trend of decreasing onset time with increasing Reynolds number, along with the deviation associated with the bifurcation observed near $Re_{\infty} = 5650$

Overall, these polynomial expressions provide compact surrogate models that approximate the results of the high-fidelity simulations across $Re_{\infty}$. Once calibrated, the regression models enable rapid prediction of the critical flow parameters without the need for repeated computationally expensive simulations, making them useful for parametric studies and data-driven analysis of rotating-cylinder wake dynamics. However, there are some challenges associated with using a rudimentary technique such as polynomial regression such as: (i) high-degree polynomials suffering from overfitting, especially near regions with sharp variations such as the bifurcation point. The model may capture noise or local fluctuations rather than the underlying physical trend which reduces its robustness when applied to new $Re_{\infty}$. (ii) Polynomial models have poor extrapolation capability, outside the range of $Re_{\infty}$ used for training \cite{hastie2009elements}. Polynomial terms grow rapidly and can produce unrealistic predictions, especially for high-degree polynomials such as those used for maximum $C_l$ and $C_d$. (iii) Numerical instability and oscillatory behavior (referred to as Runge-type oscillations \cite{burden2011numerical}) can occur with high-degree polynomial fits. These oscillations may introduce artificial wiggles in the regression curve, especially near the boundaries of the dataset.

For these reasons, while polynomial regression is useful for establishing a first-order surrogate model of the simulation results, more flexible machine-learning approaches such as ANNs or Bayesian regression can provide better generalization and stability when modeling nonlinear fluid-dynamical behavior across a wide parameter space \cite{brunton2020machine}.

\subsection{Bayesian regression applied to bifurcation analysis}

Bayesian regression provides a probabilistic framework for modelling relationships between variables by treating the model parameters as random variables rather than fixed quantities. Here, the objective is to model the dependence of key response variables like maximum $C_l$ and $C_d$, and onset time of instability on $Re_{\infty}$. Within the Bayesian framework, prior probability distributions are assigned to the regression parameters based on assumed ranges or prior physical understanding of the flow behavior. These priors are subsequently updated using Bayes’ theorem as the high-fidelity data are incorporated, resulting in posterior distributions that quantify the most probable parameter values. Unlike classical regression approaches that provide single point estimates, Bayesian regression yields posterior probability distributions for the model parameters and predictions, enabling the estimation of uncertainty bounds in the predicted aerodynamic quantities \cite{bishop2006pattern}. This capability is particularly valuable for the present dataset, where nonlinear flow phenomena such as shear-layer instabilities and the bifurcation observed near $Re_{\infty} = 5650$ can introduce variability in the relationships. The probabilistic formulation therefore allows the regression model to represent not only the mean trends in the high-fidelity data but also the associated uncertainty. Model performance can be assessed using statistical metrics such as the Bayesian information criterion (BIC) or cross-validation procedures. The effectiveness of Bayesian regression depends on the amount and quality of available data. For relatively moderate-sized datasets, such as the set of high-fidelity simulations generated in the present study, prior distributions can act as a regularization mechanism that prevents overfitting and stabilizes parameter estimation. In addition, hierarchical Bayesian formulations can improve robustness to localized fluctuations in the CFD data that may arise from complex wake interactions. When the dataset becomes sufficiently large and the error distribution is well behaved, Bayesian estimates tend to converge toward results similar to classical maximum-likelihood regression, but with the advantage of providing a probabilistic interpretation of the predictions. Despite these advantages, Bayesian regression also introduces additional computational cost, particularly when the model complexity increases or when sampling-based inference techniques such as Markov Chain Monte Carlo (MCMC) are used to estimate posterior distributions \cite{murphy2012machine}. Furthermore, the predictive accuracy of the model can be affected if the assumed prior distributions are poorly specified or if the statistical assumptions do not adequately reflect the underlying flow physics. Nevertheless, the flexibility of Bayesian inference—through methods such as MCMC sampling or variational inference—makes it a useful tool for analyzing nonlinear high-fidelity datasets where uncertainty quantification and probabilistic interpretation of the regression model are important. 

\begin{figure}[!ht]
\centering
\includegraphics[width=0.7\textwidth]{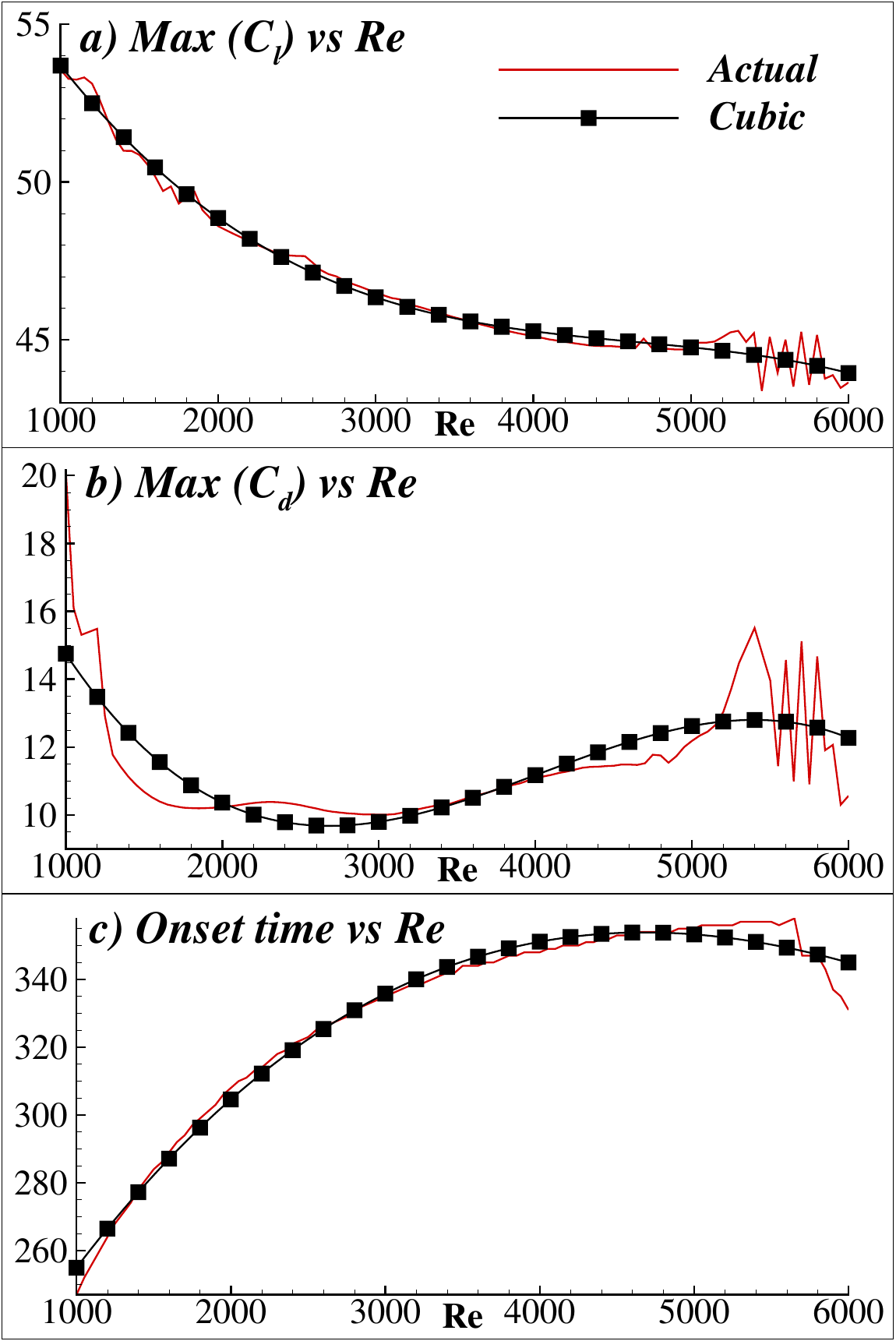}
\caption{Comparison of simulated variation of (a) maximum $C_l$, (b) maximum $C_d$, and (c) onset time of instability, as a function of $Re_{\infty}$ with Bayesian regression predictions using cubic-fitted curves.}
\label{fig9}
\end{figure}

Figure \ref{fig9} illustrates the results of Bayesian regression using cubic basis functions to model the variation of three key quantities: maximum $C_l$ and $C_d$, and the onset time of instability, with $Re_{\infty}$. The cubic basis formulation represents the response variables as linear combinations of cubic functions of $Re_{\infty}$, while the Bayesian framework treats the regression coefficients as random variables and estimates their posterior distributions from the simulation data. A primary advantage of the cubic basis formulation is its simplicity and interpretability. With only a small number of parameters, the model provides a smooth global approximation of the simulation data while maintaining a relatively low computational cost. This makes the method suitable for preliminary surrogate modeling of the relationship between $Re_{\infty}$ and the aerodynamic quantities of interest. The cubic representation provides numerical stability and reduced risk of overfitting compared with higher-degree polynomial models shown in Fig. \ref{fig8}. Because the model contains fewer coefficients, the regression is less sensitive to small fluctuations in the dataset and tends to capture the overall trend of the flow behavior rather than fitting local noise. When combined with the Bayesian formulation, the model additionally provides uncertainty quantification through posterior distributions of the regression coefficients. For parameters that exhibit relatively smooth variation with $Re_{\infty}$, such as the onset time of instability shown in Fig. \ref{fig9}(c), the cubic basis model reproduces the overall trend reasonably well. 

The aerodynamic forces in the rotating-cylinder wake exhibit strongly nonlinear behavior, particularly near the bifurcation region around $Re_{\infty} = 5650$. The cubic Bayesian model is too restrictive to capture these localized variations accurately. This limitation is most evident for maximum $C_l$ and $C_d$, where the simulated data show stronger nonlinear trends and fluctuations. Because the cubic basis functions impose a smooth global curve across the entire $Re_{\infty}$ range, the model may underpredict local peaks or fail to reproduce rapid changes associated with wake transitions. Global basis functions respond to all data points simultaneously, meaning that localized changes in the dataset can influence the regression across the entire domain. This can reduce predictive accuracy in regimes where the flow physics changes abruptly. The predictive capability of the model is evaluated using 20\%, 15\% and 10\% validation data. For maximum $C_d$, the regression accuracy increases from 57.56\% for 80\% training data to 61.38\% and 63.96\% for 85\% and 90\% training data, respectively, as summarized in Table \ref{tab5}. For maximum $C_l$, the model performs significantly better, achieving accuracies of 98.63\%, 99.13\%, and 99.48\% for the same training fractions. The onset time of instability is predicted with very high accuracies of 98.86\%, 99.14\%, and 99.00\%, respectively.

Overall, the Bayesian cubic regression model provides a stable and computationally efficient baseline surrogate model for the simulation data. It captures the broad trends in the variation of maximum $C_l$ and $C_d$, and onset of instability onset time with $Re_{\infty}$, while also providing probabilistic interpretation through Bayesian inference. However, its limited flexibility restricts its ability to represent the complex nonlinear behavior of the wake dynamics, particularly near the bifurcation region.

\begin{table}[h!]
\centering
\caption{Performance of various Bayesian regression models}
\label{tab:bayesian_regression_lines}
\begin{tabular}{|l|l|c|c|}
\hline
\textbf{Cases} & \textbf{Basis function} & \textbf{TD (\%)} & \textbf{Max Accuracy (\%)} \\
\hline
\multirow{3}{*}{Max $C_d$ vs $Re_{\infty}$}
 & Cubic    & 80/85/90 & 57.56/ 61.38/ 63.96 \\
\cline{2-4}
 & Spline   & 80/85/90 & 81.46/ 88.90/ 93.10 \\
\cline{2-4}
 & Gaussian & 80/85/90 & 45.23/ 49.76/ 51.80 \\
\hline
\multirow{3}{*}{Max $C_l$ vs $Re_{\infty}$}
 & Cubic    & 80/85/90 & 98.63/ 99.13/ 99.48 \\
\cline{2-4}
 & Spline   & 80/85/90 & 98.84/ 99.54/ 99.71 \\
\cline{2-4}
 & Gaussian & 80/85/90 & 98.80/ 99.30/ 99.68 \\
\hline
\multirow{3}{*}{Onset time vs $Re_{\infty}$}
 & Cubic    & 80/85/90 & 98.86/ 99.14/ 99.00 \\
\cline{2-4}
 & Spline   & 80/85/90 & 99.83/ 99.98/ 99.98 \\
\cline{2-4}
 & Gaussian & 80/85/90 & 99.71/ 99.87/ 99.85 \\
\hline
\end{tabular}
\label{tab5}
\end{table}

\begin{figure}[!ht]
\centering
\includegraphics[width=0.7\textwidth]{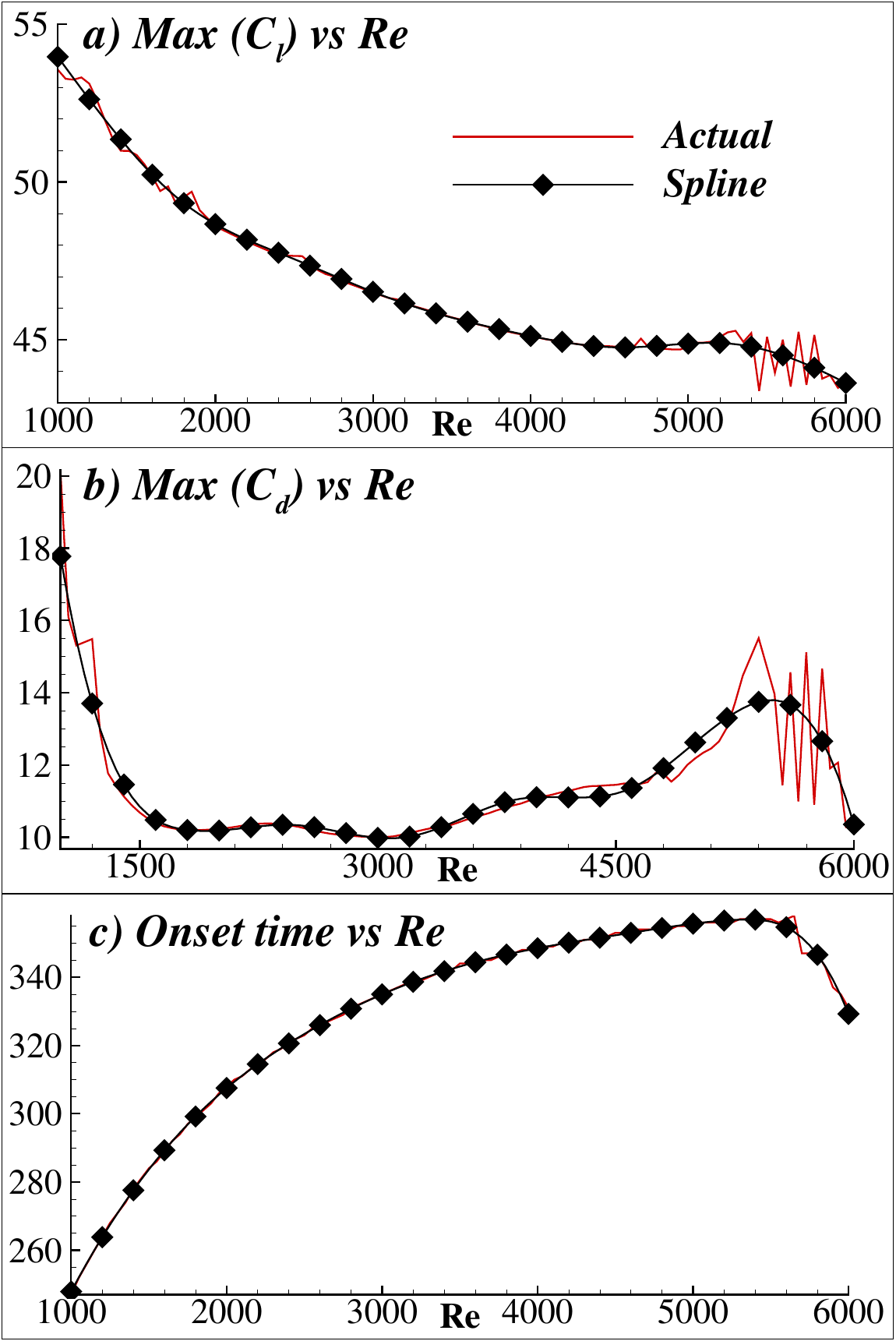}
\caption{Comparison of simulated variation of (a) maximum $C_l$, (b) maximum $C_d$, and (c) onset time of instability, as a function of $Re_{\infty}$ with Bayesian regression predictions using spline-fitted curves.}
\label{fig10}
\end{figure}

Figure \ref{fig10} presents the results of Bayesian regression using B-spline basis functions to model the dependence of the maximum $C_l$, maximum $C_d$, and onset time of instability on $Re_{\infty}$. Unlike global polynomial regression, the B-spline formulation represents the response as a piecewise polynomial function, allowing the regression model to adapt locally to variations in the simulation data while preserving smoothness across the entire $Re_{\infty}$-range. In the present implementation, cubic B-splines are employed with eight knots distributed across the investigated $Re_{\infty}$ interval. These knots partition the $Re_{\infty}$-range into segments within which the response is approximated by cubic polynomials. The resulting spline representation ensures $C^2$ continuity, providing a smooth transition between neighboring segments while allowing sufficient flexibility to capture nonlinear behavior associated with wake dynamics and the bifurcation observed at higher $Re_{\infty}$. The spline construction yields approximately eleven basis functions, forming the regression design matrix used to represent the variation of the three response variables with $Re_{\infty}$. Within the Bayesian framework, the spline coefficients are treated as random variables and assigned hierarchical prior distributions. A global shrinkage parameter ($\tau$), assumed to follow a half-normal distribution, regulates the overall magnitude of the spline coefficients and acts as a regularization mechanism that prevents overfitting. Individual spline coefficients are assigned zero-mean normal priors with variances controlled by this shrinkage parameter, thereby encouraging sparse representations while allowing the model to adapt to local variations in the simulation data.

The predictive capability of the model is evaluated using 20\%, 15\% and 10\% validation data. For maximum $C_d$, the regression accuracy increases from 81.46\% for 80\% training data to 88.90\% and 93.10\% for 85\% and 90\% training data, respectively, as summarized in Table \ref{tab5}. For maximum $C_l$, the model performs significantly better, achieving accuracies of 98.84\%, 99.54\%, and 99.71\% for the same training fractions. Similarly, the onset time of instability is predicted with very high fidelity, with accuracies of 99.83\%, 99.98\%, and 99.98\%, respectively. This approach demonstrates superior predictive performance with an accuracy typically exceeding 80\% in case of maximum $C_d$ , as the piecewise nature of splines allows the model to adapt locally to data variations while the Bayesian framework provides uncertainty quantification through posterior sampling via MCMC methods. The hierarchical prior structure combined with standardized input features ensures numerical stability during sampling, while the increased model complexity (11 parameters versus 4 in cubic polynomial) is justified by significantly improved fit quality and the model's ability to capture intricate fluid dynamics phenomena that exhibit varying behavior across different $Re_{\infty}$. 

\begin{figure}[!ht]
\centering
\includegraphics[width=0.7\textwidth]{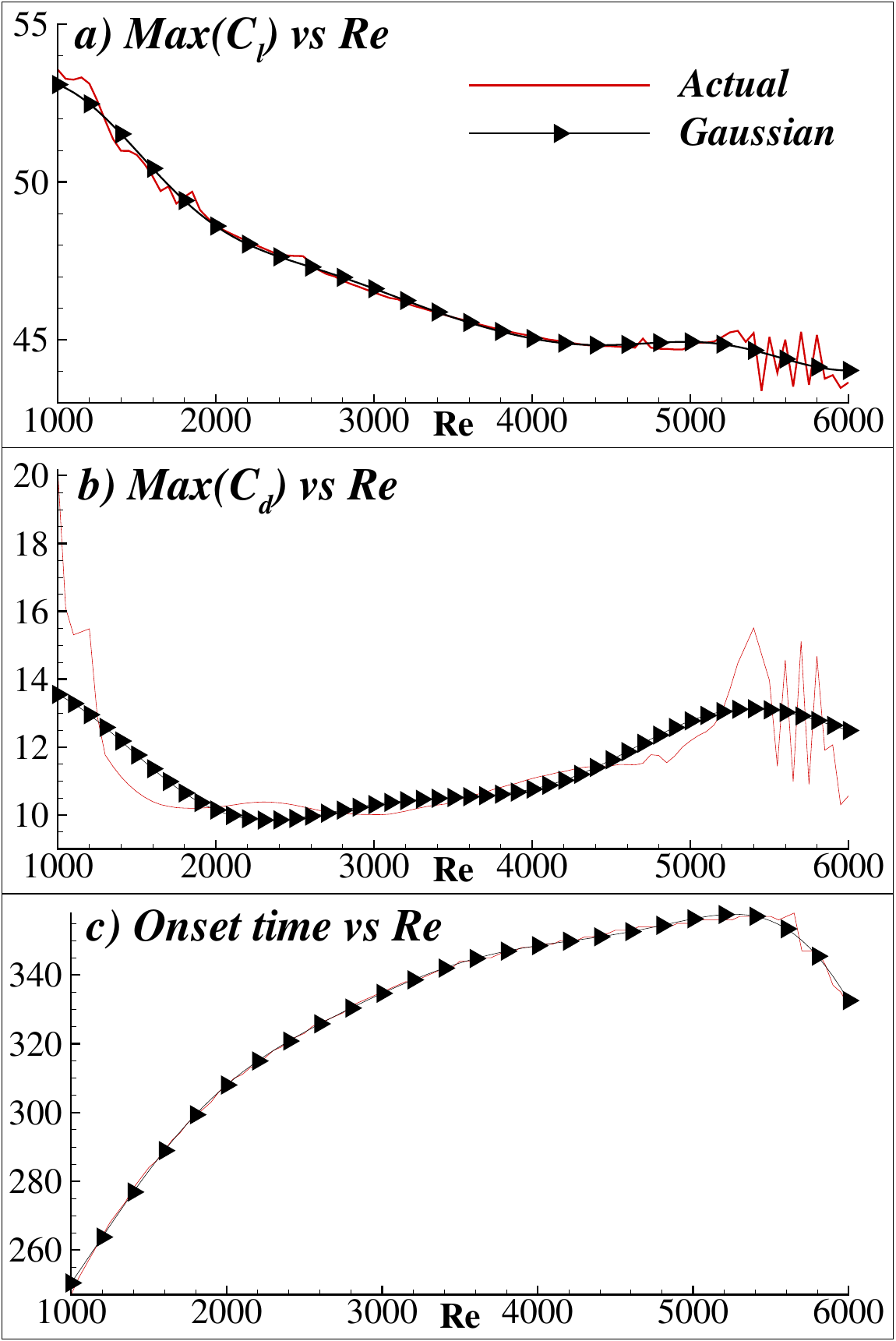}
\caption{Comparison of simulated variation of (a) maximum $C_l$, (b) maximum $C_d$, and (c) onset time of instability, as a function of $Re_{\infty}$ with Bayesian regression predictions using Gaussian-fitted curves.}
\label{fig11}
\end{figure}

Figure \ref{fig11} presents the results of a Bayesian regression model based on Gaussian radial basis functions (RBF) used to predict onset time of instability, maximum $C_l$, and maximum $C_d$ as functions of $Re_{\infty}$. In contrast to global polynomial models, the RBF formulation represents the regression function as a weighted combination of localized Gaussian kernels, allowing the model to capture nonlinear variations in the simulation data through a kernel-based approximation. In the present implementation, eight Gaussian basis functions are distributed across the investigated $Re_{\infty}$ range using evenly spaced centers. Each basis function is defined by a Gaussian kernel with a fixed width parameter ($\sigma = 0.5$), which determines the spatial extent over which a basis function influences the prediction. The Gaussian basis function takes the form \cite{buhmann2000radial} \[
\phi(x) = \exp\!\left(-\tfrac{1}{2} \left(\frac{x-c}{\sigma}\right)^{2}\right)\] \noindent where $c$ denotes the center of the kernel. This formulation allows each kernel to represent localized variations in the response variables, enabling the model to adapt to different dynamical regimes across $Re_{\infty}$ domain. Within the Bayesian framework, the weights associated with the Gaussian basis functions and the intercept term are assigned normal prior distributions. The regression model therefore learns the optimal linear combination of Gaussian kernels that best represents the relationship between $Re_{\infty}$ and simulated response variables. Posterior distributions of the model parameters are estimated through Markov Chain Monte Carlo (MCMC) sampling, performed with 2000 iterations and 1000 tuning steps, which also provides uncertainty quantification for the predicted quantities.

For maximum $C_d$, the prediction accuracy remains comparatively low, reaching 45.23\%, 49.76\%, and 51.80\% for 80\%, 85\%, and 90\% training data, respectively, as summarized in Table \ref{tab5}. This reduced accuracy indicates that the drag response exhibits sharper variations and more complex behavior with $Re_{\infty}$, which are not fully captured by the chosen kernel configuration. In contrast, the model performs significantly better for maximum $C_l$, achieving accuracies of 98.80\%, 99.30\%, and 99.68\% for the corresponding training fractions. Similarly, the onset time of instability is predicted with very high accuracy, reaching 99.71\%, 99.87\%, and 99.85\%, respectively.

Overall, the Gaussian RBF model demonstrates excellent predictive capability for quantities that vary smoothly with $Re_{\infty}$, such as onset time of instability and maximum $C_l$. The localized nature of the Gaussian kernels allows the model to capture gradual transitions in the flow behavior across different $Re_{\infty}$ regimes. However, the approach is less effective in representing the more irregular variations observed in $C_d$, particularly near the bifurcation point where abrupt changes in wake dynamics occur. Despite this limitation, the combination of kernel-based representation and Bayesian inference provides a flexible modeling framework that offers both predictive capability and uncertainty quantification for the simulation dataset.

\subsection{Modeling bifurcation with an artificial neural network}

An artificial neural network (ANN) is a data-driven computational framework composed of interconnected layers of neurons, including an input layer, multiple hidden layers, and an output layer \cite{brunton2020machine}. Each neuron applies a weighted transformation to the incoming signals, followed by a nonlinear activation, allowing the network to learn complex input–output relationships \cite{goodfellow2016deep}. In the context of the present high-fidelity dataset for compressible flow past a rapidly rotating cylinder, the ANN is employed to model the nonlinear dependence of key response variables: maximum $C_l$, maximum $C_d$, and onset time of instability, on $Re_{\infty}$. The dataset used for training the ANN consists of 101 simulation cases between $Re_{\infty} = 1000$ and 6000, in intervals of 50, for which total computational cost is approximately one million core-hours. Such a dataset captures intricate flow physics, including shear-layer instabilities, vortex interactions, and bifurcation behavior, making it well-suited for data-driven modelling approaches. The architecture of the ANN is designed to balance model complexity and generalization capability. The final configuration consists of a deep neural network with 12 hidden layers, containing 256, 256, 128, 128, 64, 64, 32, 32, 16, 16, 8, and 8 neurons, respectively. This progressively decreasing structure enables the network to first learn high-dimensional feature representations and then compress them into more compact forms, improving stability and reducing overfitting. Preliminary experiments with shallower networks (4, 6, and 8 layers) were unable to adequately capture the nonlinear trends in the simulation data, while deeper architectures (e.g., 14 layers) resulted in reduced accuracy, likely due to over-parameterization and training difficulties. The 12-layer configuration was therefore selected as the optimal compromise. To model the strong nonlinearities present in the dataset, the exponential linear unit (ELU) activation function is employed. ELU is particularly effective for such problems because it maintains nonlinearity while mitigating vanishing gradient issues during backpropagation \cite{goodfellow2016deep}, thereby improving convergence for deep architectures. For training, the network uses the Adam optimizer with a default learning rate, ensuring efficient gradient-based optimization, while the Huber loss function is adopted due to its robustness to outliers that may arise from localized flow variations or numerical uncertainties in the simulation data. 

Depending on the input $Re_{\infty}$ , the functional relationship with the three critical parameters may follow a linear, quadratic, cubic, logarithmic or a reciprocal function. This choice is made on the basis of the flow physics inferred from simulation plots. For example, linear function is adopted when there is direct relationship between $Re_{\infty}$  and the output flow parameter. The quadratic term, on the other hand, is chosen when the secondary instabilities in the transitional flow dominate and often exhibit quadratic growth. Similarly, when compressibility effects are important in the flow, a cubic distribution is often adopted. The logarithmic term is used to satisfy log-law velocity profile in turbulent boundary layer and reciprocal function is used to provide viscous correction. This enhances the input space and capture complex nonlinear relationships between $Re_{\infty}$ and critical flow parameters such as onset time of instability, maximum $C_l$  and maximum $C_d$ . To design the ANN models, linear, quadratic and cubic distributions are considered for all three output parameters, while for maximum $C_l$ versus $Re_{\infty}$, we use an additional logarithmic function. Similarly, for maximum $C_d$ versus $Re_{\infty}$, an additional reciprocal function is used. To emphasize the model’s predictive capability in high $Re_{\infty}$-regimes, custom sample weights are assigned, enhancing the contribution of data samples in the critical parts of the distributions where fluctuations are significant.

The ANN is trained using randomly selected data sets from the 101 2D simulations to prevent overfitting. Four training scenarios were tested using 82\%, 85\%, 87\% and 90\% of the data. Data sets were split into three subsets: training set, testing set and validation set respectively to mitigate the risk of overfitting of the model. To refine the ANN model, we tested the model using 20\%, 15\% and 10\% validation data. For the maximum $C_d$ versus $Re_{\infty}$ distribution, the batch size is chosen as 20. The maximum accuracy for 82\%, 85\%, 87\% and 90\% training data is obtained as 78.08\%, 81.52\%, 85.88\% and 90.03\%, respectively as shown in Table \ref{tab6}. For the maximum $C_l$ versus $Re_{\infty}$ distribution, the batch size is chosen as 15. The maximum accuracy for 82\%, 85\%, 87\% and 90\% training data is obtained as 99.2\%, 99.83\%, 99.8\% and 99.8\%, respectively as shown in Table \ref{tab7}. For $C_l$, the results do not vary much between the various training datasets used, which indicates that its variation with $Re_{\infty}$ is relatively straightforward to model compared to $C_d$ . This is affirmed by the gradual dip in the $C_l$ distribution, wherein small-amplitude fluctuations appear for $Re_{\infty} > 5650$. The fluctuations in the $C_d$ distribution are much steeper than that for $C_l$, which makes it harder to model using ANN. For the onset time of instability versus $Re_{\infty}$ distribution, the ANN model uses a batch size of 10. The maximum accuracy for 82\%, 85\%, 87\% and 90\% training data is obtained as 99.84\%, 99.94\%, 99.98\% and 99.97\%, respectively as shown in Table \ref{tab8}. The variation of onset time with $Re_{\infty}$ follows a simple functional relationship, for which the ANN results does not vary much between the four training data sets. This suggests that the level of refinement offered by the 101 data sets is not necessary for $C_l$ and onset time of instability, but for $C_d$  there are significant variations with $Re_{\infty}$, a dense data set is paramount to the success of the ANN model.

\begin{table}
\begin{center}
\def~{\hphantom{0}}
\caption{Performance of the ANN trained with maximum $C_d$ vs $Re_{\infty}$ distribution.}
\begin{tabular}{|c|c|c|c|c|}
 \hline
 Training Data & Epochs & MAE & MSE & Accuracy(\%) \\ [3pt]
 \hline
 
 {\bf 82\%} & {\bf 100} & {\bf 0.4610} & {\bf 1.2516} & {\bf 78.08} \\ \hline
 82\% & 200 & 0.4983 & 1.4807 & 74.06 \\ \hline
 82\% & 300 & 0.4861 & 1.4509 & 74.59 \\ \hline
 82\% & 400 & 0.5544 & 1.4972 & 73.78 \\ \hline
 82\% & 500 & 0.5721 & 1.2865 & 77.47 \\ \hline
 82\% & 550 & 0.4938 & 1.2931 & 77.35 \\ \hline
 82\% & 600 & 0.4918 & 1.3826 & 75.78 \\ \hline
 85\% & 100 & 0.4616 & 1.5711 & 76.37 \\
 \hline
 85\% & 200 & 0.4188 & 1.2914 & 80.58 \\
 \hline
 85\% & 300 & 0.4924 & 1.4693 & 77.90 \\
 \hline
 {\bf 85\%} & {\bf 400} & {\bf 0.4919} & {\bf 1.2286} & {\bf 81.52} \\
 \hline
 85\% & 500 & 0.6252 & 2.0898 & 68.57 \\
 \hline
 85\% & 600 & 0.4963 & 1.9580 & 70.55 \\
 \hline
 87\% & 100 & 0.4247 & 1.2578 & 80.71 \\ \hline
 {\bf 87\%} & {\bf 200} & {\bf 0.4927} & {\bf 0.9209} & {\bf 85.88} \\ \hline
 87\% & 300 & 0.5901 & 1.7100 & 73.77 \\ \hline
 87\% & 400 & 0.4570 & 1.2986 & 80.08 \\ \hline
 87\% & 500 & 0.3956 & 1.1764 & 81.96 \\ \hline
 87\% & 600 & 0.5707 & 1.4556 & 77.68 \\ \hline
 87\% & 650 & 0.4250 & 1.0106 & 84.50 \\ \hline
 87\% & 700 & 0.5153 & 1.4583 & 77.63 \\ \hline
 90\% & 100 & 0.4692 & 1.6481 & 79.29 \\
 \hline
 90\% & 200 & 0.4421 & 1.6197 & 79.65 \\
 \hline
 90\% & 300 & 0.4959 & 1.9651 & 75.31 \\
 \hline
 90\% & 400 & 0.4145 & 1.4783 & 81.42 \\
 \hline
 90\% & 500 & 0.5187 & 1.2285 & 84.56 \\ 
 \hline
  \textbf{90\%} & \textbf{600} & \textbf{0.2892} & \textbf{0.7937} & \textbf{90.03} \\ \hline
 \end{tabular}
\label{tab6}
\end{center}
\end{table}

\begin{table}
\begin{center}
\def~{\hphantom{0}}
\caption{Performance of the ANN trained with maximum $C_l$ vs $Re_{\infty}$ distribution.}
\begin{tabular}{|c|c|c|c|c|}
 \hline
 Training Data & Epochs & MAE & MSE & Accuracy(\%) \\ [3pt]
 \hline
 
 82\% & 100 & 0.1807 & 0.0838 & 99.02 \\ \hline
 82\% & 200 & 0.1854 & 0.0899 & 98.95 \\ \hline
 82\% & 300 & 0.1637 & 0.0721 & 99.15 \\ \hline
 82\% & 400 & 0.1421 & 0.0740 & 99.13 \\ \hline
 82\% & 500 & 0.1571 & 0.0787 & 99.08 \\ \hline
 82\% & 600 & 0.1279 & 0.0741 & 99.13 \\ \hline
 {\bf 82\%} & {\bf 650} & {\bf 0.1263} & {\bf 0.6620} & {\bf 99.22} \\ \hline
 85\% & 100 & 0.1131 & 0.0257 & 99.69 \\
 \hline
  \textbf{85\%} & \textbf{200} & \textbf{0.0835} & \textbf{0.0139} & \textbf{99.83} \\
 \hline
 85\% & 300 & 0.0780 & 0.0182 & 99.78 \\
 \hline
 85\% & 400 & 0.0969 & 0.0225 & 99.73 \\
 \hline
 85\% & 500 & 0.0856 & 0.0156 & 99.81 \\
 \hline
 85\% & 600 & 0.1058 & 0.0256 & 99.69 \\
 \hline
 87\% & 100 & 0.1100 & 0.0314 & 99.52 \\ \hline
 87\% & 200 & 0.0802 & 0.0169 & 99.74 \\ \hline
 {\bf 87\%} & {\bf 250} & {\bf 0.0718} & {\bf 0.0131} & {\bf 99.80} \\ \hline
 87\% & 300 & 0.0815 & 0.1361 & 99.72 \\ \hline
 87\% & 400 & 0.0226 & 0.1504 & 99.66 \\ \hline
 87\% & 500 & 0.0824 & 0.0178 & 99.73 \\ \hline
 87\% & 600 & 0.0971 & 0.0216 & 99.67 \\ \hline
 87\% & 700 & 0.1150 & 0.0224 & 99.66 \\ \hline
 90\% & 100 & 0.0977 & 0.0316 & 99.57 \\
 \hline
 90\% & 200 & 0.1368 & 0.0253 & 99.65 \\
 \hline
 90\% & 300 & 0.1163 & 0.0337 & 99.54 \\
 \hline
 {\bf 90\%} & {\bf 400} & {\bf 0.0838} & {\bf 0.0146} & {\bf 99.80} \\
 \hline
 90\% & 500 & 0.0916 & 0.0154 & 99.79 \\
 \hline
 90\% & 600 & 0.0734 & 0.0156 & 99.79 \\ \hline
 \end{tabular}
\label{tab7}
\end{center}
\end{table}

\begin{table}[h!]
\centering
\caption{Performance of the ANN trained with onset time vs $Re_{\infty}$ distribution.}
\begin{tabular}{|c|c|c|c|c|}
 \hline
 Training Data & Epochs & MAE & MSE & Accuracy(\%) \\ [3pt]
 \hline
 
 82\% & 100 & 0.9121 & 2.2974 & 99.78 \\ \hline
 {\bf 82\%} & {\bf 200} & {\bf 0.5973} & {\bf 1.6889} & {\bf 99.84} \\ \hline
 82\% & 300 & 0.7538 & 2.5002 & 99.76 \\ \hline
 82\% & 400 & 0.7377 & 2.2118 & 99.79 \\ \hline
 82\% & 500 & 0.9619 & 3.2419 & 99.69 \\ \hline
 82\% & 600 & 1.3888 & 5.0565 & 99.52 \\ \hline
 85\% & 100 & 1.1892 & 1.7795 & 99.82 \\
 \hline
 85\% & 200 & 1.0162 & 1.5759 & 99.84 \\
 \hline
 85\% & 300 & 1.4377 & 2.5794 & 99.74 \\
 \hline
 85\% & 400 & 0.5879 & 0.6527 & 99.93 \\
 \hline
 {\bf 85\%} & {\bf 500} & {\bf 0.6053} & {\bf 0.5754} & {\bf 99.94} \\
 \hline
 85\% & 600 & 0.6216 & 0.5596 & 99.94 \\
 \hline
 87\% & 100 & 0.4311 & 0.3009 & 99.96 \\ \hline
 87\% & 200 & 0.3460 & 0.2290 & 99.97 \\ \hline
 87\% & 300 & 0.8004 & 0.9251 & 99.89 \\ \hline
 87\% & 400 & 0.6843 & 0.8809 & 99.90 \\ \hline
 {\bf 87\%} & {\bf 500} & {\bf 0.3475} & {\bf 0.4315} & {\bf 99.98} \\ \hline
 87\% & 600 & 0.3393 & 0.2009 & 99.98 \\ \hline
 87\% & 700 & 0.2974 & 0.1351 & 99.98 \\ \hline
 90\% & 100 & 0.5451 & 0.4412 & 99.95 \\
 \hline
 90\% & 200 & 0.4723 & 0.3443 & 99.96 \\
 \hline
 90\% & 300 & 0.5218 & 0.4133 & 99.96 \\
 \hline
 90\% & 400 & 0.6363 & 0.6311 & 99.94 \\
 \hline
 90\% & 500 & 0.6213 & 0.5543 & 99.94 \\
 \hline
  \textbf{90\%} & \textbf{600} & \textbf{0.4551} & \textbf{0.2574} & \textbf{99.97} \\ \hline
  
 \end{tabular}
\label{tab8}
\end{table}

\begin{figure}[!ht]
\centering
\includegraphics[width=\textwidth]{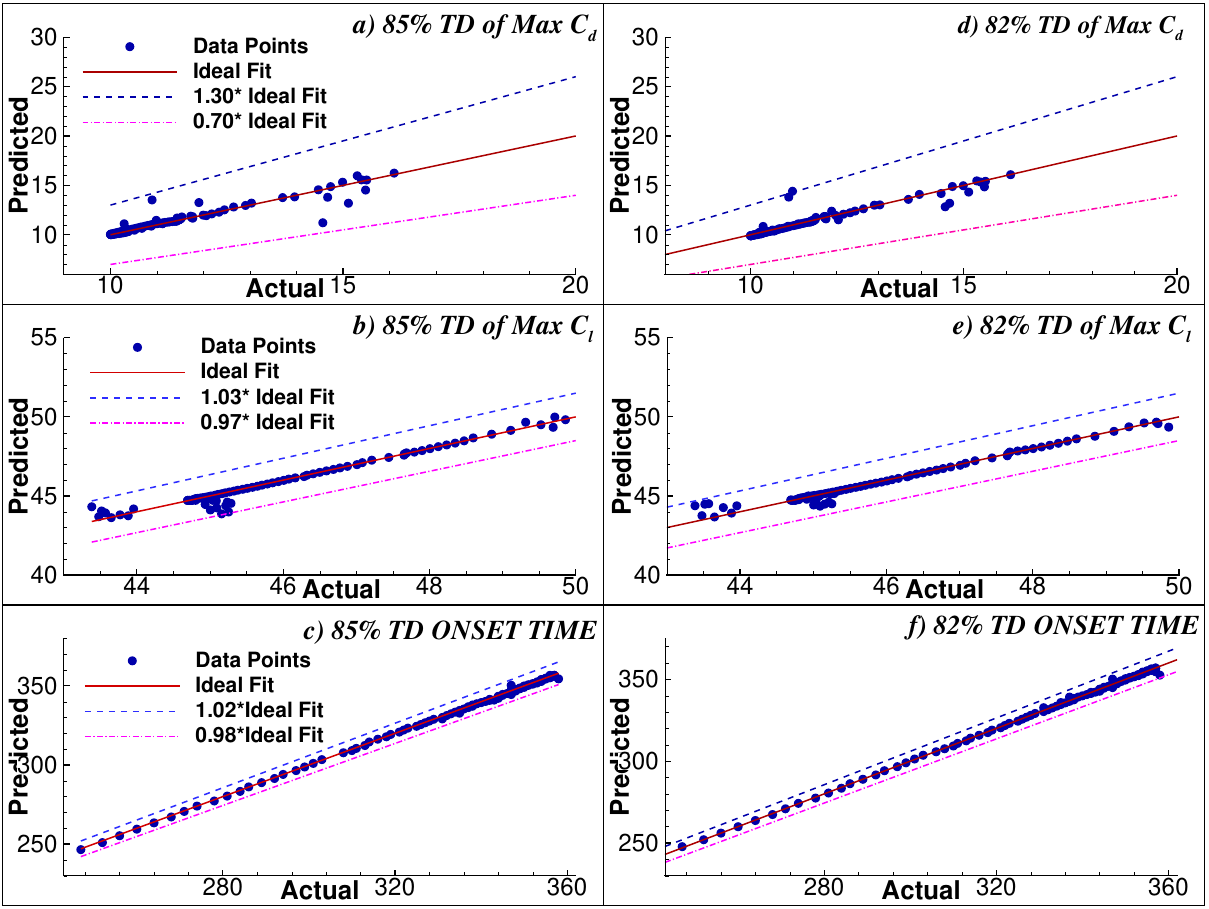}
\caption{Parity plot of predicted value from ANN and actual value from simulation for (a)-(c) 85\% training data, and (d)-(f) 82\% training data. The maximum $C_d$, maximum $C_l$, and onset time are shown in frames (a), (d); (b), (e); and (c),(f), respectively.}
\label{fig12}
\end{figure}

Figure \ref{fig12} presents parity plots comparing the ANN-predicted values with the corresponding simulation data for maximum $C_d$, maximum $C_l$, and the onset time of instability, using two different training splits (85\% and 82\%). The diagonal line in each subplot represents the ideal agreement between prediction and simulation, while the bands around it indicate acceptable deviation margins. For the 85\% training dataset shown in Figs. \ref{fig12}(a)-(c), the ANN predictions show strong agreement with the simulation data for all three quantities. The predictions for maximum $C_l$ and onset time are tightly clustered along the ideal line, with only minimal scatter, indicating excellent predictive accuracy and very low error. The narrow deviation bands (approximately $\pm 3\%$ for maximum $C_l$ and $\pm 2\%$ for onset time) further confirm that the ANN effectively captures the underlying nonlinear relationship between $Re_{\infty}$ and these quantities. In contrast, maximum $C_d$ exhibits a comparatively larger spread of data points around the ideal line. Although the overall trend is still captured, the predictions deviate more significantly, as reflected by the wider tolerance band (approximately $\pm 30\%$). This behavior is consistent with earlier observations that maximum $C_d$ shows stronger nonlinear fluctuations and sensitivity to wake dynamics, making it more challenging for ANN to model accurately. For the 82\% training dataset shown in Figs. \ref{fig12}(d)-(f), a similar trend is observed, but with a slight degradation in predictive performance. The scatter around the ideal line increases marginally for all three quantities, particularly for $C_d$. However, predictions for maximum $C_l$ and onset time remain closely aligned with the ideal fit, demonstrating that ANN performs well even with reduced training data.

\begin{figure}[!ht]
\centering
\includegraphics[width=\textwidth]{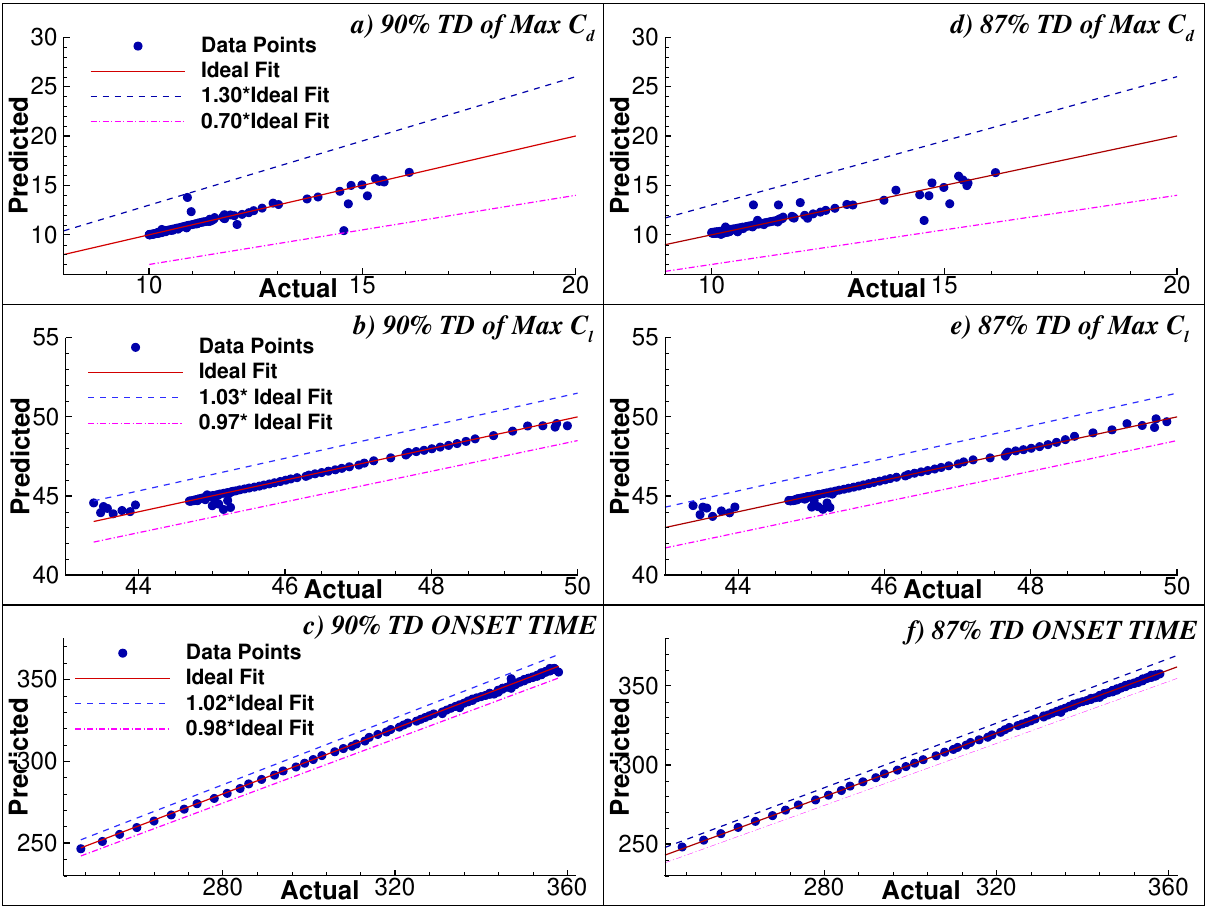}
\caption{{Parity plot of predicted value from ANN and actual value from simulation for (a)-(c) 90\% training data, and (d)-(f) 87\% training data. The maximum $C_d$, maximum $C_l$, and onset time are shown in frames (a), (d); (b), (e); and (c),(f), respectively.}}
\label{fig13}
\end{figure}

Figure \ref{fig13} presents parity plots comparing ANN predictions with simulation data for maximum $C_d$, maximum $C_l$, and onset time of instability, for two higher training fractions: 90\% and 87\%. For the 90\% training dataset depicted in Figs. \ref{fig13}(a)-(c), the ANN exhibits excellent predictive accuracy across all three quantities. The predictions for maximum $C_l$ and onset time are almost perfectly aligned with the ideal fit line, with extremely tight clustering and very narrow deviation bounds. This indicates that the ANN has effectively learned the smooth and deterministic dependence of these parameters on $Re_{\infty}$. For maximum $C_d$, although the ANN captures the overall trend better than in Fig. \ref{fig12}, the scatter is visibly larger compared to the other two quantities. This increased difficulty in modeling drag is due to its sensitivity to wake unsteadiness, vortex interactions, and post-bifurcation dynamics. For the 87\% training dataset shown in Figs. \ref{fig13}(d)-(f), the trends remain largely consistent, with only a slight degradation in accuracy. 

\subsection{Generalizing beyond the data: ANN-based generative model for bifurcation parameters}

While the ANN developed thus far has demonstrated strong performance as a regression tool on the available 101 high-fidelity simulation datasets, its true utility lies in its ability to generalize beyond the discrete training points and act as a surrogate generator of flow physics in previously unexplored regions of the parameter space. This capability is particularly important near critical bifurcation regimes, where small changes in $Re_{\infty}$ can lead to disproportionately large variations in flow response. To assess this, the ANN is now employed as a generative model, tasked with predicting flow behavior at finer resolutions than those used during training. The focus is placed on the distribution of maximum $C_d$ vs $Re_{\infty}$, as $C_d$ has consistently exhibited the highest sensitivity, strongest nonlinear fluctuations, and greatest prediction difficulty, especially in the vicinity of the bifurcation threshold ($Re_{\infty} \approx 5650$). A hierarchical refinement strategy is adopted:

\noindent {\bf Level-1 refinement:}
The ANN is trained using the original 101 datasets with a uniform spacing of $\Delta Re_{\infty} = 50$. It is then used to generate intermediate predictions in the range $5300 \le Re_{\infty} \le 6000$ at a finer resolution of $\Delta Re_{\infty} = 25$. The predicted values are subsequently validated against independently computed high-fidelity simulations, serving as the ground truth. This step evaluates the model’s ability to interpolate within a highly nonlinear and transitional regime.

\noindent {\bf Level-2 refinement:}
The training dataset is further enriched by incorporating the Level-1 data (i.e., $\Delta Re_{\infty} = 25$ in the range $5300 \le Re_{\infty} \le 6000$), while retaining the coarser spacing elsewhere. The ANN is then used to predict an even finer distribution with $\Delta Re_{\infty} = 12$ in the same critical range. These predictions are again benchmarked against high-fidelity simulations to assess accuracy.

This two-level refinement framework enables a systematic evaluation of the ANN’s multi-resolution generalization capability. In particular, it examines whether the model can (i) reconstruct missing intermediate states, (ii) capture sharp gradients and fluctuations associated with post-bifurcation dynamics, and (iii) maintain predictive fidelity as the resolution of the parameter space increases. By focusing on the most challenging observable ($C_d$), this approach establishes a stringent test for the ANN’s robustness. Successful performance would demonstrate that the trained network is not merely interpolating known data, but is effectively learning the underlying physics-informed mapping, thereby enabling reliable generation of high-resolution parametric trends at a fraction of the computational cost of direct simulations. For refinement Level-1 and Level-2, we use the same ANN architectures as before but in two stages: in first stage, ANN acts as a global estimator. It captures the general \lq \lq shape" of the curve. In second stage, ANN acts as a correction model. By feeding stage-1's prediction back into stage-2, the second model can focus specifically on the error or the refinement needed to match the data, especially in the sensitive $Re_{\infty}$ regime. For this, we have used sample weights for $Re_{\infty} > 5300$  to force the model to prioritize accuracy in the highly fluctuating regime. 

The Level-1 refinement results provide a stringent test of the ANN’s generative capability in the most sensitive regime ($5300 \le Re_{\infty} \le 6000$), where 
maximum $C_d$ exhibits strong nonlinear fluctuations due to proximity to the bifurcation point. A key observation from Table \ref{tab9} is that the ANN is able to capture the overall magnitude and trend of maximum $C_d$ across the entire range. The predicted values consistently lie within the correct order of magnitude and broadly follow the envelope of variation seen in the simulated data. This indicates that the network has successfully learned the global mapping between 
$Re_{\infty}$ and drag response, even when queried at unseen intermediate points. However, a closer inspection reveals important limitations: (i) in regions where the simulated maximum $C_d$ varies smoothly (e.g., $Re_{\infty}$ = 5475, 5775, 5825, 5975), the ANN predictions show good agreement, with relatively small deviations. This suggests that the model performs well in quasi-linear or weakly nonlinear regimes. (ii) In contrast, near strong fluctuation zones (e.g., $Re_{\infty}$ = 5425, 5525, 5575, 5625), the ANN exhibits significant discrepancies, often overpredicting or underpredicting the drag. For example, at $Re_{\infty}$ = 5425, a sharp drop in simulated $C_d$ is not captured by the ANN and at $Re_{\infty}$ = 5525 and 5625, the ANN fails to reproduce low drag values, instead predicting much higher values. These deviations highlight a fundamental challenge, i.e., the ANN tends to produce a smoothed representation of the response, thereby damping high-frequency variations associated with multi-mode interactions and post-bifurcation dynamics. This behavior is typical of data-driven models trained on relatively coarse sampling, where fine-scale oscillations are under-resolved in the training data. From a physical standpoint, the discrepancies arise because the region near $Re_{\infty} = 5650$ is characterized by mode competition, intermittent transitions, and nonlinear coupling, leading to rapid, non-monotonic changes in drag. Such localized features are difficult to reconstruct without sufficiently dense training data.

\begin{table}[h!]
\centering
\caption{Comparison of simulated and predicted values of maximum $C_d$ versus $Re_{\infty}$ for refinement level-1.}
\label{tab:re_maxcd_updated}
\begin{tabular}{c c c}
\hline
\textbf{$Re_{\infty}$} & \textbf{Simulated} & \textbf{Predicted} \\
\hline
5325 & 15.88907 & 15.561416 \\
5375 & 16.20292 & 15.503049 \\
5425 & 13.01796 & 15.298844 \\
5475 & 11.33620 & 11.355942 \\
5525 & 9.39166  & 14.019061 \\
5575 & 13.14121 & 11.364592 \\
5625 & 8.91177  & 11.370447 \\
5675 & 12.99026 & 14.694981 \\
5725 & 16.26261 & 13.497245 \\
5775 & 9.29724  & 10.911540 \\
5825 & 10.37776 & 11.774801 \\
5875 & 12.86799 & 11.988627 \\
5925 & 11.32557 & 12.015352 \\
5975 & 9.76156  & 10.585518 \\
\hline
\end{tabular}
\label{tab9}
\end{table}

\begin{figure}[!ht]
\centering
\includegraphics[width=\textwidth]{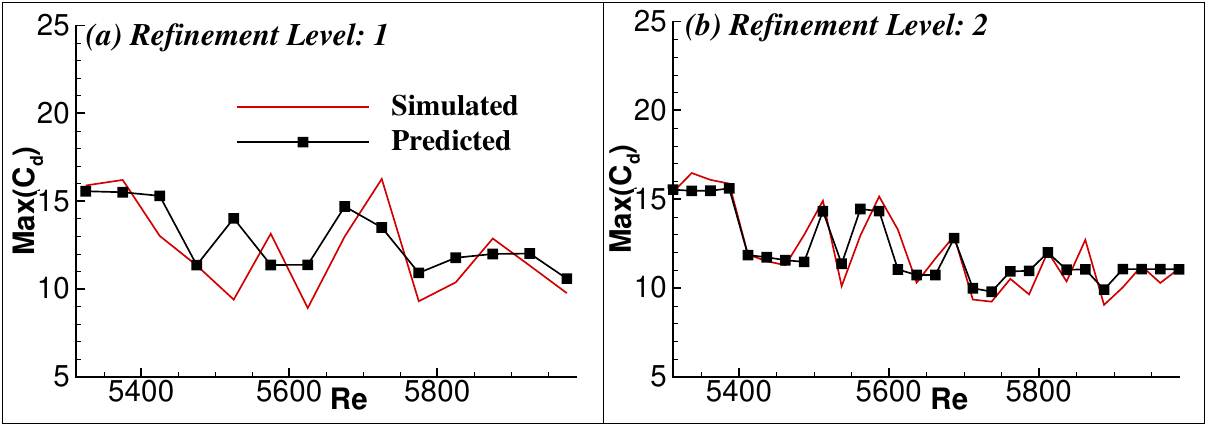}
\caption{Comparison of simulated maximum $C_d$ vs $Re_{\infty}$ with the predicted values from the generative ANN for (a) level-1 refinement and (b) level-2 refinement.}
\label{fig14}
\end{figure}

The ANN for Level-1 refinement shown in Fig. \ref{fig14}(a) demonstrates strong interpolation capability at a global level, successfully predicting intermediate values between training points. The model shows reduced accuracy in highly nonlinear, fluctuation-dominated regions, particularly near the bifurcation threshold. Predictions exhibit a smoothing bias, indicating that the ANN captures the mean trend but not sharp local extrema. These results justify the need for Level-2 refinement, where additional training data at finer resolution can help the ANN better resolve localized physics.

Table \ref{tab10} for Level-2 refinement represents a more demanding assessment of the ANN’s generative capability, where the model is trained on a partially enriched dataset ($\Delta Re_{\infty} = 50$ globally and $\Delta Re_{\infty} = 25$ near bifurcation) and is then used to predict finer-scale variations at $\Delta Re_{\infty} = 12$ within $5300 \le Re_{\infty} \le 6000$. A clear improvement over Level-1 refinement is immediately evident. The ANN predictions now show much closer agreement with the simulated values across most of the $Re_{\infty}$ range. The model is able to track both high and low values of maximum $C_d$ more accurately, capture the envelope of oscillations, and reduce large deviations that were previously observed near the bifurcation region. This indicates that the inclusion of intermediate training points has enhanced the ANN’s ability to resolve nonlinear behavior. For smooth / weakly nonlinear regions (e.g., $Re_{\infty} \approx 5412-5462, 5812-5987$), predictions are highly accurate, with very small deviations. The ANN effectively reproduces both the magnitude and trend of maximum $C_d$, demonstrating strong interpolation capability. In moderately fluctuating regions (e.g., $Re_{\infty} \approx$ 5487–5587, 5687–5762), ANN captures the trend and relative variation reasonably well, although minor overprediction or underprediction persists. Unlike Level-1 refinement, however, the model no longer completely misses local behavior. In highly nonlinear / bifurcation-dominated regions (around $Re_{\infty} \approx$ 5600–5660, some discrepancies remain such as, the ANN underpredicting a relatively high simulated value at $Re_{\infty} = 5612$. However, compared to Level-1, the magnitude of error is reduced, and the ANN better reflects the presence of local extrema.

\begin{table}[h!]
\centering
\caption{Comparison of simulated and predicted values of maximum $C_d$ versus $Re_{\infty}$ for refinement level-2.}
\label{tab:re_maxcd}
\begin{tabular}{c c c}
\hline
\textbf{$Re_{\infty}$} & \textbf{Simulated} & \textbf{Predicted} \\
\hline
5312 & 15.39288 & 15.535071 \\
5337 & 16.47013 & 15.469234 \\
5362 & 16.08484 & 15.472721 \\
5387 & 15.87401 & 15.606017 \\
5412 & 11.88417 & 11.849112 \\
5437 & 11.50252 & 11.729534 \\
5462 & 11.28333 & 11.561792 \\
5487 & 12.99211 & 11.467979 \\
5512 & 14.90838 & 14.320218 \\
5537 & 10.10479 & 11.358833 \\
5562 & 12.95313 & 14.443304 \\
5587 & 15.14851 & 14.323092 \\
5612 & 13.28426 & 11.041621 \\
5637 & 10.29789 & 10.728555 \\
5662 & 11.65725 & 10.733514 \\
5687 & 12.92219 & 12.807559 \\
5712 & 9.34612  & 9.9857496 \\
5737 & 9.22807  & 9.7897220 \\
5762 & 10.51033 & 10.931631 \\
5787 & 9.63669  & 10.963257 \\
5812 & 12.03401 & 11.994020 \\
5837 & 10.36712 & 11.024311 \\
5862 & 12.70932 & 11.044918 \\
5887 & 9.04714  & 9.9056274 \\
5912 & 10.03349 & 11.061749 \\
5937 & 11.23440 & 11.062416 \\
5962 & 10.28342 & 11.058087 \\
5987 & 11.08938 & 11.046937 \\
\hline
\end{tabular}
\label{tab10}
\end{table}

The ANN for Level-2 refinement shown in Fig. \ref{fig14}(b) demonstrates reduced smoothing bias, indicating improved sensitivity to local fluctuations. Predictions now better align with local peaks and troughs, rather than only capturing the global trend. The model demonstrates enhanced ability to represent multi-mode oscillatory behavior in the post-bifurcation regime. The improved performance can be directly linked to the increased resolution of training data in the critical region. By exposing the ANN to intermediate states ($\Delta Re_{\infty} = 25$), the model gains access to local variations in wake dynamics, rapid transitions associated with mode competition, and non-monotonic changes in drag arising from nonlinear vortex interactions, as depicted in Fig. \ref{fig8}.

Level-2 refinement demonstrates that the ANN is not merely interpolating but is progressively becoming a physics-aware generative surrogate, capable of reconstructing fine-scale variations in a highly nonlinear regime. While small discrepancies persist near sharp transitions, the overall predictive fidelity is significantly improved, confirming that data enrichment in critical regions is essential for accurate generative modeling of bifurcation-driven flows.

\section{Summary and Conclusions}
This study presents a comprehensive investigation of compressible flow past a rapidly rotating cylinder over a wide Reynolds number range ($Re_{\infty} = 1000-6000$), combining high-fidelity numerical simulations with data-driven modeling approaches. The analysis focuses on the evolution of aerodynamic loads, spectral characteristics of unsteady forces, and the onset of instability, with particular emphasis on the nonlinear dynamics emerging near the critical bifurcation regime. The flow is shown to transition from periodic vortex shedding at lower Reynolds numbers to multi-mode oscillatory behavior \cite{joshi2023exploring} at higher Reynolds numbers. Spectral analysis of the drag coefficient reveals the presence of multiple dominant frequencies ($P_1$–$P_3$), whose relative amplitudes and spacing evolve with $Re_{\infty}$, indicating increasing nonlinear mode interactions and harmonic generation. A critical bifurcation is identified near $Re_{\infty} \approx 5650$, beyond which the flow exhibits strong intermittency, amplitude modulation, and enhanced sensitivity to small parameter variations. This is further reflected in the non-monotonic variation of maximum lift and drag coefficients, as well as a clear departure of the instability onset time from a smooth parabolic trend.

To model these complex dependencies, several regression frameworks are evaluated. Polynomial regression, while capable of capturing global trends, is shown to suffer from overfitting and poor extrapolation in regions with strong nonlinear fluctuations. Bayesian regression approaches provide improved flexibility and uncertainty quantification, with B-spline basis functions demonstrating superior capability in representing localized nonlinear behavior compared to Gaussian radial basis functions, particularly in regimes with sharp transitions.

Artificial neural networks (ANNs) are then developed as high-capacity surrogate models trained on 101 high-fidelity simulation datasets. The ANN demonstrates excellent predictive performance for maximum lift coefficient and instability onset time, with near-perfect agreement with simulation data. For maximum drag coefficient, which is highly sensitive to wake dynamics and post-bifurcation effects, the ANN captures the overall trend but exhibits reduced accuracy in regions dominated by strong fluctuations and mode competition.

Beyond regression, the ANN is further assessed as a generative surrogate model to predict flow behavior at unseen Reynolds numbers. A hierarchical refinement strategy is employed to evaluate its generalization capability. In Level-1 refinement, the ANN trained on coarsely spaced data ($\Delta Re_{\infty} = 50$) successfully reconstructs intermediate states ($\Delta Re_{\infty} = 25$), capturing global trends but smoothing out localized extrema. In Level-2 refinement, the inclusion of additional training data ($\Delta Re_{\infty} = 25$) significantly improves predictive fidelity, enabling the ANN to better resolve fine-scale nonlinear variations and partially recover local peaks and troughs in the drag response. These results demonstrate that the ANN progressively transitions from a regression tool to a physics-informed generative model, with accuracy strongly dependent on the resolution of training data in critical regimes. 

Overall, the study establishes that data-driven models, when trained on high-fidelity CFD datasets, can effectively capture the underlying nonlinear mapping between flow parameters and aerodynamic response. The ANN-based framework, in particular, offers a computationally efficient alternative for reconstructing complex flow behavior, especially in parameter regimes where direct simulations are prohibitively expensive.

While the present work demonstrates the potential of ANN-based surrogate and generative modeling, several directions remain for further improvement and extension such as incorporating governing equations or physical constraints (e.g., through physics-informed neural networks). This could improve predictive accuracy in highly nonlinear regimes and reduce reliance on dense training data. Similarly, integration of Bayesian neural networks or ensemble methods could enable rigorous quantification of predictive uncertainty, particularly near bifurcation points where sensitivity is highest. Furthermore, extending the methodology to other bluff-body configurations or rotating geometries would help assess the universality and robustness of the proposed approach.

\section*{Credit authorship contribution statement}
{\bf Sanjeev Kumar:} Visualization, Methodology, Investigation, Formal analysis, Data curation, Writing - Reviewing and Editing. {\bf Santosh Kumar:} Visualization, Methodology, Investigation, Writing - Reviewing and Editing. {\bf Aditi Sengupta:} Conceptualization and Supervision, Data curation, Formal analysis, Investigation, Writing- Original draft preparation. 

\section*{Declaration of competing interest}
The authors declare that they have no known competing financial interests or personal relationships that could have appeared to influence the work reported in this paper.

\section*{Data Availability}
\noindent The data that support the findings of this study are available from Aditi Sengupta (aditi@iitism.ac.in) upon reasonable request. The Python programs used for polynomial, Bayesian regression and ANN are available as supplementary data.

\section*{Acknowledgements}
\noindent The authors acknowledge the use of the HPC facility CHANDRASEKHAR provided by IIT (ISM) Dhanbad for all the computations reported here.

\bibliographystyle{elsarticle-num} 
 \bibliography{rot-cyl}

\end{document}